\documentclass[%
 reprint,
 amsmath,amssymb,
 aps,
 prl,
 floatfix,
]{revtex4-2}

\usepackage{physics}
\usepackage{graphicx}
\usepackage{bm} % bold math
\usepackage[hyperfootnotes=true,linkcolor=blue,menucolor=black,colorlinks=false]{hyperref} % add hypertext capabilities
\usepackage[capitalise]{cleveref}
\usepackage[acronym]{glossaries}
\glsdisablehyper
\usepackage{dsfont}
\usepackage[normalem]{ulem}

\newacronym{qm}{QM}{Quantum Mechanics}
\newacronym{povm}{POVM}{Positive Operator Valued Measure}
\newacronym{ann}{ANN}{Artificial neural network}
\newacronym{hnn}{HNN}{Hopfield neural network}
\newacronym{am}{AM}{associative memory}

\newcommand{\ccite}[1]{Ref.~\cite{#1}}
\newcommand{\disp}[1]{\mathcal{D}\left[#1\right]}
\newcommand{\lv}{\mathcal{L}}
\newcommand{\hmt}{\hat{\mathrm{H}}}
\newcommand{\Id}{\mathds{I}}
\newcommand{\opa}{\hat{a}}
\newcommand{\opad}{\hat{a}^\dagger}

\newcommand{\inpar}[1]{\left( #1 \right)}
\newcommand{\insqr}[1]{\left[ #1 \right]}
\newcommand{\inbrc}[1]{\left\lbrace #1 \right\rbrace}
\begin{document}

\title{Quantum Associative Memory with a Single Driven-Dissipative Nonlinear Oscillator}

\author{Adrià Labay-Mora}
\email{alabay@ifisc.uib-csic.es}
\author{Roberta Zambrini}
\author{Gian Luca Giorgi}%
\affiliation{%
 Institute for Cross Disciplinary Physics and Complex Systems (IFISC) UIB-CSIC, Campus Universitat Illes Balears, Palma de Mallorca, Spain.
}%

\date{\today}
\begin{abstract}
    Algorithms for associative memory typically rely on a network of many connected units. The prototypical example is the Hopfield model, whose generalizations to the quantum realm are mainly based on open quantum Ising models. We propose a realization of associative memory with a single driven-dissipative quantum oscillator exploiting its infinite degrees of freedom in phase space. The model can improve the storage capacity of discrete neuron-based systems in a large regime and we prove successful state discrimination between $n$ coherent states, which represent the stored patterns of the system. These can be tuned continuously by modifying the driving strength, constituting a modified learning rule. We show that the associative-memory capability is inherently related to the existence of a spectral separation in the Liouvillian superoperator, which results in a long timescale separation in the dynamics corresponding to a metastable phase.
\end{abstract}

\maketitle

\glspl{ann} are brain-inspired computational systems that can solve and model numerous kinds of tasks, ranging from pattern and speech recognition \cite{amari1998gradient,bishop2006pattern} to big data analysis \cite{hinton}. An important family of \glspl{ann} is given by attractor networks, whose temporal evolution settles on stable solutions, exploited in a wide range of problems \cite{amit_1989,hertz2018introduction} with the prominent example of \gls{am}. In an \gls{am} task, a system stores a set of memory states. Then, it is interrogated using a clue state similar but not necessarily identical to one of the memories; a system equipped with \gls{am} can identify the stored pattern most similar to the clue according to a properly defined distance. \glspl{am} are commonly modeled through the (classical) \gls{hnn} \cite{hopfield1982neural}, which makes use of a network of binary neurons, and exhibits stable attractors -- the memories -- defined through a proper \textit{learning rule} written in the weights of the neural connections \cite{amit_1989,hertz2018introduction}. One main limitation of the \gls{hnn} is that the number of patterns that can be stored is much smaller than the dimension of the network itself \cite{amit1985storing,amit1987statistical}.   

Quantum machine learning aims to find ways to exploit the features of quantum mechanics for machine learning purposes \cite{wittek2014quantum,biamonte2017quantum,Dunjko_2018,RevModPhys.91.045002}. In the context of quantum \gls{am}, generalizations of classical models are mainly based on the quantized version of the \gls{hnn} \cite{Gopalakrishnan,diamantini2006quantum,PhysRevLett.107.277201,PhysRevA.95.032310,rotondo2018open,PhysRevA.98.042308,carollo2021exactness,fiorelli2021phase,enhancing2021marsh}, where binary systems are replaced by quantum spins, and where the necessary dissipative dynamics are provided by the interaction with some external bath (which can also encode the learning rule \cite{PhysRevLett.114.143601,fiorelli2020signatures}). The main findings concern the existence of dynamical phases, not found in classical systems, that can be employed in new types of retrieval. Yet, memories remain strings of classical bits. Still, an open point is the promise that the richer dynamics of quantum systems can improve the storage capacity, that is, the number of memories over the system size. A general discussion about the possibility of achieving such a quantum advantage can be found in \cite{sanpera2021capacity,bodeker2022optimal}, where the storage capacity is estimated according to the Gardner program \cite{gardner1988space,gardner1988optimal}. However, direct application to specific models does not seem to give conclusive answers~\cite{gratsea2021storage,benatti2022pattern}.

In this Letter, we take an alternative route to \gls{am} in quantum systems moving from spin networks to a single driven-dissipative nonlinear quantum oscillator where one can exploit its (in principle infinite) number of degrees of freedom. The main ingredient of our approach lies in the nonlinearity which determines the form and phase symmetry of the steady state, changing from (almost classical) coherent states to purely quantum states, depending on the model parameters. Together with a metastable dynamical phase long enough compared to all timescales relevant to pattern recognition and memory retrieval. Concerning \glspl{am}, metastability allows systems that converge towards a unique steady state to span a manifold of relevant addressable memories \cite{brinkman2022metastable}.

In principle, a quantum oscillator spans an infinite Hilbert space with potentially unlimited storage capacity \cite{sanpera2021capacity}. This can be seen as a (generally complex) network whose computational nodes can be built using every orthogonal basis of the Liouville space (a similar approach was taken in \ccite{govia2021quantum} in the context of quantum reservoir computing). Nevertheless, we are bounded by the size of the metastable manifold. Considering the minimum Hilbert space size needed to correctly describe the system dynamics, we will show that our model can achieve a higher storage capacity than the (classical and quantum) discrete neuron models. 

Let us briefly review the concept of metastability, which emerges whenever disparate timescales are present in the evolution of a dynamical system \cite{brinkman2022metastable}. In our case, as we will see, metastability can be traced back to the presence of a separation in the Liouvillian spectrum \cite{minganti2018spectral,macieszczak2016towards} and is in close connection with quantum entrainment and dissipative phase transitions \cite{cabot2021metastable}. It is characterized by the long-lived occupation of high Liouvillian modes and is normally observed after a short transient time and before the final relaxation towards the steady state.
% in our case, as we will see, the spectral decomposition can...

For a system described by the Gorini-Kossakowski-Sudarshan-Lindblad master equation $\partial_t \rho = \lv \rho$ \cite{lindblad1976generators,gorini1976completely}, the dynamics can be understood in terms of the set of complex eigenvalues $\inbrc{\lambda_j}$ of the (non-Hermitian) Liouvillian superoperator $\lv$ and of the right ($\inbrc{R_j}$) and left ($\inbrc{L_j}$) eigenvectors, obeying, respectively, $\lv R_j = \lambda_j R_j$ and $\lv^\dagger L_j = \lambda_j^* L_j$ with normalization $\tr L_j^\dagger R_k = \delta_{jk}$ \cite{minganti2018spectral}. Then, assuming the presence of at least one steady state $\rho_{ss}$ (which is always true in finite dimensions \cite{evans1977generators,baumgartner2008analysis}), the time evolution of a state $\rho(0)$ can be decomposed as
\begin{equation} \label{eq:pp_modes_decomp}
    \rho(t) = \rho_{ss} + \sum_{j > 1} \tr[L_j^\dagger \rho(0)] e^{\lambda_j t} R_j\ ,
\end{equation}
where for convenience the eigenvalues are sorted such that $0 \ge \Re\lambda_j \ge \Re\lambda_{j + 1}$.
% where we assumed the presence of a single steady state which survives at $t \to \infty$.
   
A metastable dynamical phase will emerge before the final relaxation whenever there is a large separation between two consecutive eigenvalues, i.e. $\tau_n \gg \tau_{n + 1}$ where $\tau_n^{-1} = -\Re \lambda_n$ \cite{macieszczak2016towards}. This divides the decay into different timescales: a fast regime for $t < \tau_{n + 1}$, a metastable period where dynamics are apparently frozen for $\tau_{n + 1} < t < \tau_{n}$; and finally, the last decay for $t > \tau_{n}$. In the middle region, the dynamics can be approximated by $\rho(t) = \sum_{l=1}^n p_l(t) \mu_l$ \cite{macieszczak2021theory}, where $\inbrc{ \mu_l }_{l=1}^n$ are the metastable states spanning the metastable manifold \cite{SuppMaterial} and $\inbrc{p_l(t)}_{l=1}^n$ are quasiprobabilities, as they might take negative values, but satisfy that their sum is $1$.

Our quantum model for \gls{am} consists of a driven-dissipative oscillator described by the master equation
\begin{equation} \label{eq:pp_me_nm}
    \pdv{\rho}{t} = -i [\hmt_n, \rho] + \gamma_1 \disp{\opa}\rho + \gamma_m \disp{\opa^m} \rho\ ,
\end{equation}
where we have standard terms for linear (single-photon) and nonlinear (multiphoton) damping \cite{mundhada2017four,gevorkyan1999coherent} with rates $\gamma_1$ and $\gamma_m$ respectively. The Hamiltonian, which contains a $n$-order squeezing drive \cite{braunstein1987generalized,lang2021multi}, in the rotation frame and after the parametric approximation is
\begin{equation} \label{eq:pp_ham_nm}
    \hmt_n = \Delta \opad \opa + i \eta \left[\opa^n e^{i \theta n} - (\opad)^n e^{-i \theta n} \right]\ .
\end{equation}
Here, $\Delta = \omega_0 - \omega_s$ is the detuning between the natural oscillator frequency and that of the squeezing force, $\eta$ and $\theta$ the magnitude and phase of the driving, respectively. We observe that the model possesses $\mathbb{Z}_n$ symmetry, that is, the transformation $\opa \to \opa \exp(i2\pi/n)$ leaves the master equation invariant \cite{minganti2023dissipative}.

Although particular solutions have been found for specific cases \cite{bartolo2016exact,cabot2021metastable}, no general analytical solution exists for \cref{eq:pp_me_nm}. We can restrict to the case $m = n$ and write it as \citep{mundhada2017four,mirrahimi2014dynamically,ma2021quantum}
\begin{equation} \label{eq:pp_me_n}
    \pdv{\rho}{t} = -i\Delta [\opad\opa, \rho] + \gamma_1 \disp{\opa}\rho + \gamma_n \disp{\opa^n - \beta^n} \rho\ ,
\end{equation}
where $\beta^n = 2 \eta e^{i \theta n}/\gamma_n$ corresponds to the amplitude of $n$ symmetrically distributed coherent states or \emph{lobes}
\begin{equation} \label{eq:pp_lobes_coh}
    \ket{\beta_j} = \ket*{\beta e^{i(2 j + 1) \pi / n}},\qquad j=1,\dots,n
\end{equation}
which span the kernel of the nonlinear damping term in \cref{eq:pp_me_n}. We notice that $\beta$ is a function of the ratio between squeezing strength and nonlinear damping. In the limit of small detuning and large $\beta$, we observe numerically that the lobes become almost orthogonal ($F(\beta) = \abs{\braket{\beta_j}{\beta_{j+1\bmod n}}}^2 \to 0$), and thus the steady state can be well approximated by $\rho_{ss} \approx (1/n) \sum_{j= 1}^n \op{\beta_j}$ \cite{gevorgyan2008superposition}. Instead, in the absence of squeezing in \cref{eq:pp_ham_nm}, only a single solution with $\beta = 0$ persists. In the following, we fix $\theta = 0$ and $\Delta = 0.4\gamma_1$.
    
\begin{figure}
    \centering
    \includegraphics[width=\linewidth]{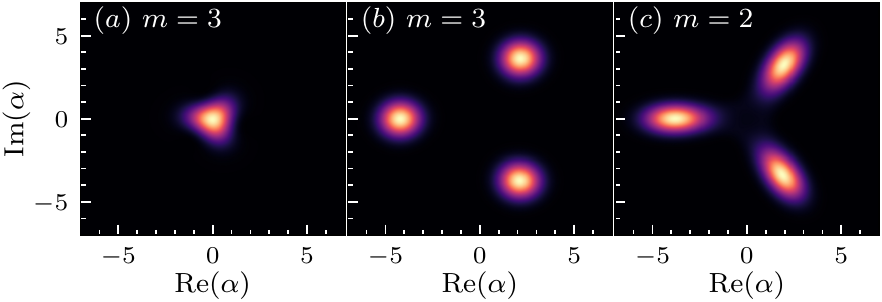}
    \caption{Wigner representation of the steady state for $n=3$ (normalized). Parameters: (a) $\gamma_2/\gamma_1 = 0.2$, $\eta/\gamma_1 = 0.1$; (b) $\gamma_2/\gamma_1 = 1.5$, $\eta/\gamma_1 = 2.7$; (c) $\gamma_2/\gamma_1 = 0.2$, $\eta/\gamma_1 = 1.455$.}
    \label{fig:lobes4}
\end{figure}

By numerically solving the steady state equation $\lv \rho_{ss} = 0$, we show in \cref{fig:lobes4} its Wigner representation for four different parameter choices \cite{numerics}. In the first row, we can see two different situations for $n = m = 3$: in panel (a) we have set $\eta \ll \Delta$, which makes the lobes indistinguishable, while for $\eta > \Delta$ [panel (b)] we can appreciate three coherent states corresponding to an amplitude $\beta \sim 3$. The separation between these two regimes could also be observed at the mean-field level, as explicitly discussed in \cite[sec. S1]{SuppMaterial}. Finally, in \cref{fig:lobes4}(c), we show the steady state for $n = 3$ and $m = 2$. Here, we again see three lobes, as expected from the symmetry of the system, but now show signatures of squeezing and quantumness. This also applies to other values of $n \neq m$. In all situations, the Wigner representation is non-negative as a consequence of the linear damping which removes the coherences between states \cite{gevorgyan2008superposition,gilles1994generation}.
    
\begin{figure}
    \centering
    \includegraphics[width=\linewidth]{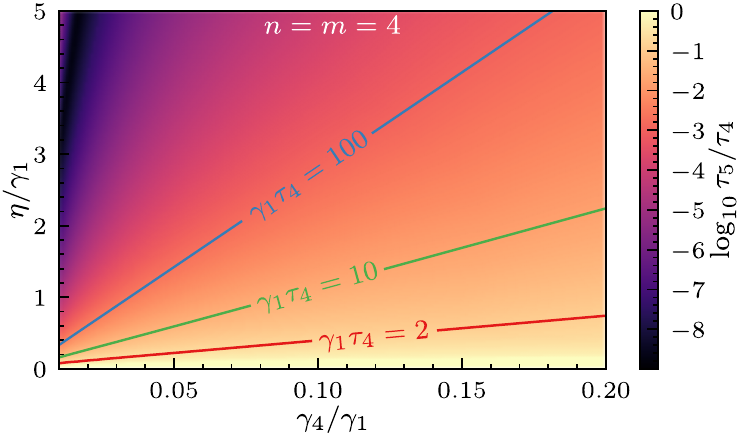}
    \caption{Separation between the $4$th and $5$th low-lying eigenvalues where darker colors indicate longer metastability. Three contour lines show the decay time of the fourth eigenvalue, the end of metastability.}
    \label{fig:phase_4}
\end{figure}
    
To establish the existence of metastability, let us explore how the separation of the Liouvillian eigenvalues depends on the system parameters. An example is given in \cref{fig:phase_4} (for $n = m = 4$). There, the separation appears between the fourth and fifth eigenvalues, which separates slow metastable dynamics from fast decay modes \cite{macieszczak2016towards}. During the slow phase, the dynamics can be approximated by $n$ metastable phases $\{ \mu_j \}_{j=1}^n$, constructed as extreme superpositions of the first $n$ eigenmodes \cite[sec. S3]{SuppMaterial}. These, in the regime of large $\beta$, are approximately equal to the coherent states in \cref{eq:pp_lobes_coh}. The larger the separation, the farther apart the lobes are, increasing the metastable properties \footnote{This is in correspondence with the results found in \ccite{sonar2018squeezing} where squeezing reduces the effects of noise and enhances the stability of the lobes.}.

The results above are consistent with particular situations studied in the literature. Concretely, the case $n = m = 2$ was studied in \ccite{cabot2021metastable} using linear amplification instead of linear damping (the presence of both damping and amplification was analyzed in \ccite{PhysRevResearch.2.033422} in the context of quantum synchronization). The change, motivated by its experimental feasibility \cite{mundhada2017four,svensson2018period,sonar2018squeezing}, leads to a slight increase in metastability because there is no competition between dissipative terms. Yet, no other qualitative difference is appreciated.

We now turn our attention to the dynamic properties that lead to the \gls{am} capabilities of the system. Our goal is to exploit the metastable dynamics to discriminate between the $n$ metastable phases. This can be seen as a generalized discrimination problem between $n$ symmetrical coherent states \cite{nair_symmetric_2012,izumi_quantum_2013} because the initial state does not have to be any of the lobes. More specifically, within the metastable transient, an initial state will move towards the closest lobe (representing one of the stored memories) and remain there for a long time. Consequently, by measuring the state within this regime, we can extract information about the corresponding lobe. Furthermore, the ability to tweak the target states using the (tunable) Liouvillian parameters can be interpreted as a modified learning rule, commonly given in \gls{ann} by changing the network weights to select the desired family of steady states \cite{hopfield1982neural}. 
    
\begin{figure}
    \centering
    \includegraphics[width=\linewidth]{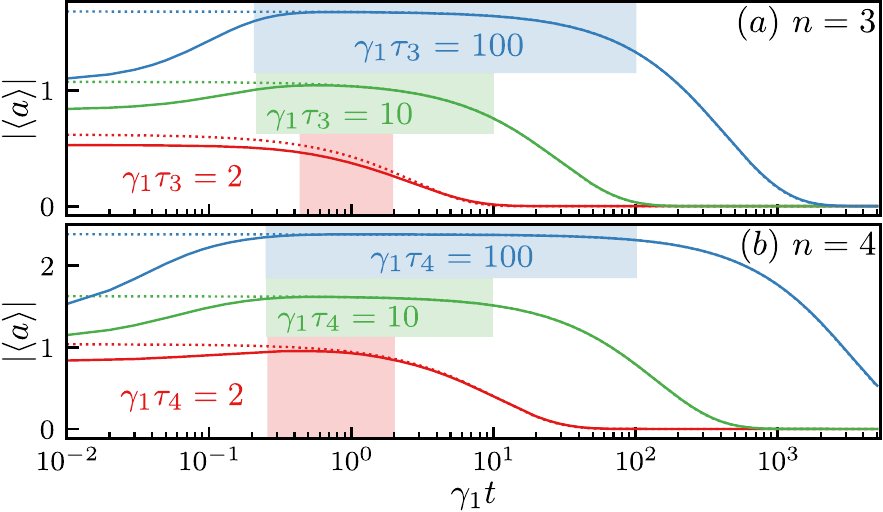}
    \caption{Time evolution of $\abs{\ev{\opa}}$ for three different parameter sets as shown in \cref{fig:phase_4}. Full master equation evolution in solid lines (for truncated Hilbert space with $\dim \mathcal{H} = 50$) and deterministic evolution in the metastable manifold in dotted lines. Shaded areas indicate the metastability regime per color.}
    \label{fig:traj}
\end{figure}

In \cref{fig:traj}, we compute the time evolution of $\ev{\opa}$ for three different values of the parameters $(\gamma_n, \eta)$ corresponding to $\gamma_1\tau_n = \inbrc{2, 10, 100}$. The initial state is a coherent state with amplitude $0.5 \beta(\gamma_n, \eta) \exp(i 2\pi /9)$, different from any lobe. Then, the evolution is evaluated by comparing the full master equation (solid lines) with the metastable approximation described in \ccite{macieszczak2021theory} (dotted lines), which, of course, is expected to be valid since the metastable transient. 
    
Looking at the upper two lines, for both $n$, we can distinguish the different dynamical regimes. First, a fast decay of the high modes (i.e. $R_{j>n}$) occurs, which takes the state from its initial amplitude to that of the lobes in a time $\tau_{n+1}$. Here, as expected, the metastable approximation fails to describe the dynamics. Then, the solution penetrates into the metastable transient where a plateau of constant amplitude is observed. From this point onwards, the two descriptions coincide with high accuracy, showing that the state is confined to the metastable manifold. Thus, in this setting, metastability is completely described by the Liouvillian spectrum. In contrast, when the separation between lobes is small (lower red lines), the metastable transient disappears. We can also appreciate a longer plateau for $n = 4$ than for $n = 3$, even after the metastable transient, which is a consequence of the slowest eigenvalues distribution~\cite[sec. S4]{SuppMaterial}.

% To quantify the efficiency of the \gls{am}, we evaluate numerically the probability that at a time $t$ a state is found at the target lobe. For that, we initialize the system in a coherent state with random amplitude $[0, 2\beta]$ and phase, and compute its Monte Carlo evolution under \cref{eq:pp_me_n}. The system is then measured with the (ambiguous) POVM $\{ P_k \}_{k=1}^n$, obtained numerically from the Liouvillian left eigenmodes \cite{macieszczak2021theory}, where each operator corresponds to a division of the phase space centered around each lobe \eqref{eq:pp_lobes_coh} \cite{lorch2019quantum}. Hence, assuming the initial state shared most similarity with the $k$-th lobe (according to the trace distance), the success probability is equal to the click probability of the $k$-th operator \cite[sec. S5]{SuppMaterial}. Finally, we average over $400$ random initial states to obtain the solid lines in \cref{fig:memory}(a).

\begin{figure}
    \centering
    \includegraphics[width=\linewidth]{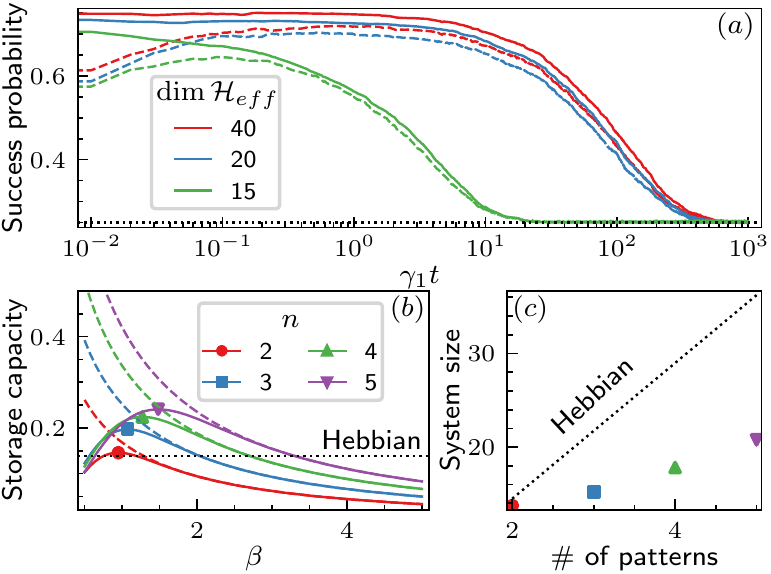}
    \caption{(a) Probability of identifying the correct lobe with time using the ambiguous (solid lines) and unambiguous (dashed lines) strategies for increasing system sizes. Parameters corresponding to the green line in \cref{fig:phase_4}b. (b) Storage capacity over lobe amplitude for a different number of patterns: $\tilde{\alpha}_c$ (solid lines) and $\alpha_c$ (dashed lines). (c) Points corresponding to the maximal storage capacity in (b). Compared with Hebbian critical capacity $\alpha_c^{Hebb} = 0.138$ (dotted line).}
    \label{fig:memory}
\end{figure}

Next, we assess the \gls{am} efficiency by numerically computing the probability that the system is found in the target lobe at each time $t$. We use a Monte Carlo simulation with a coherent state of random amplitude $[0, 2\beta]$ and phase as initial state. The system is then measured with the (ambiguous) POVM $\{ P_k \}_{k=1}^n$, obtained numerically from the Liouvillian left eigenmodes \cite{macieszczak2021theory}, where each operator corresponds to a division of the phase space centered around each lobe \eqref{eq:pp_lobes_coh} \cite{lorch2019quantum}. Hence, the success probability is equal to the click probability of the $k$th operator assuming the initial state is most similar to the $k$th lobe (according to trace distance) \cite[sec. S5]{SuppMaterial}. We repeat this process $400$ times with different initial states and average the results to obtain the solid lines in \cref{fig:memory}(a).

Focusing on the $\dim \mathcal{H}_{eff} = 40$ (solid red) line, the time evolution can be compared to the metastable evolution in \cref{fig:traj} with a plateau of high success probability that spans times even before the metastable regime begins. This is because any state in the basin of attraction of the lobe will trigger the associated operator, failing to determine whether the state has converged to the exact pattern. Thus, we repeat the calculation with a second (unambiguous) POVM, used experimentally for $m$-ary phase-shifted keys \cite{izumi2012displacement,becerra2013experimental}, that only triggers when the state is inside the metastable manifold. As a result, looking at the dashed lines in \cref{fig:memory}(a), we note that the success probability is initially small -- the state is not over any lobe -- but converges to the plateau by the start of the metastable transient, thus showing its ability to optimally discriminate the patterns in this regime.

A fundamental question in the context of \gls{am} concerns the storage capacity $\alpha_c$ of a system. While our model has an infinitely dimensional Hilbert space, the coherent-state solutions discussed so far can be described with high accuracy by truncating above high Fock state occupancy of the boson mode \cite{lang2021multi}. This allows us to adapt the definition of the storage capacity of finite systems. In \cref{fig:memory}(a), we show the effect of truncation on the lobe identification. As expected, under a certain system size, the dynamical state cannot be well approximated and metastability is lost, which leads to a fast decrease in the success probability. The dimension of the truncated Hilbert space $ \mathcal{H}_{\rm{eff}}$ represents the effective system size to be compared with the number of stored memories. Assuming that $n$ patterns can be successfully stored, one can define the storage capacity as $\alpha_c = n / \dim \mathcal{H}_{\rm{eff}}$. However, the possibility to distinguish them can be strongly hindered depending on the parameter choice \footnote{For instance, as we saw in \cref{fig:lobes4}(a), a one lobe solution is present with $n=3$ with $\beta$ small, implying a small effective dimension and consequently, a high storage capacity \cite[sec. S6]{SuppMaterial}.}. This highlights the importance of accounting for correlations between patterns \cite{gardner1988space,benatti2022pattern}. Hence, we define $\tilde{\alpha}_c= [1 - F(\beta)] \alpha_c$, with $F(\beta)$ as specified above, which vanishes for indistinguishable lobes ($\beta \to 0$) and large dimensions ($\beta \to \infty$) but is maximal for intermediate amplitudes. In \cref{fig:memory}(b), we plot the storage capacity as a function of the lobe amplitude for different values of $n$. Although contrasting different learning rules is not immediate, we compare it with the standard Hebbian rule which has been found to limit the capacity in both classical \cite{amit1985storing} and quantum \cite{enhancing2021marsh}. In this way, we can appreciate a wide range of solutions where such classical limit is exceeded \footnote{We note that the Hebbian critical capacity is found in the limit of infinite dimension (neurons) and zero temperature while in our case the dimension is finite once the Hilbert space is truncated. Nevertheless, the critical value seems to apply to finite resources too. For instance, \citet{enhancing2021marsh} found numerically that this limit persists with at least $200$ neurons, and practical examples never overcome this limit \cite{fuchs1988pattern,fiorelli2020signatures}. \label{fn:hebbianlimit}}. Further, in \cref{fig:memory}(c), we show how the maximal storage capacity of our model reduces the system size required to store the same number of patterns in a Hebbian-based \gls{hnn}.

In this Letter, we have proposed a different approach to \gls{am} considering a single driven-dissipative quantum nonlinear oscillator. We have shown that it allows for successful state discrimination during the metastable regime. Our approach shares some features with the classical continuum space limit of the Wilson-Cowan model \cite{wilson1972excitatory}, whose stochastic versions \cite{PhysRevE.75.051919,PhysRevE.82.051903} account for metastable neural population activity. In this sense, the Wigner function plays the role of a neural field whose excitations represent the stored patterns \cite[sec. S7]{SuppMaterial}. In contrast to these models where the solutions settle at long times, our \gls{am} is transient, which may provide a speed-up in the convergence towards the patterns \cite{fiorelli2019quantum}.

Even if bosonic models can potentially encode an infinite number of memories \cite{segura2000associative}, our system is upper-bounded by two values: the power of the nonlinear term $n$ and the overlap between the lobes $F(\beta)$. The latter is similar to the conditions for patterns in \glspl{hnn}, which require them to be orthogonal. At the same time, the former determines the dimension of the metastable manifold, i.e. the number of metastable solutions. In this respect, we saturate the maximum number of patterns of the system \cite{sanpera2021capacity}, and most importantly, $n$ is not upper-bounded in theory.

We can compare our proposal, where the number of solutions can be increased with the nonlinearity degree $n$, with the standard Hebbian learning strategy, where one needs to increase the dimension of the Hilbert space (number of spins). As the former can be well approximated by truncation, we have found a superior storage capacity $\tilde{\alpha}_c > \alpha_c^{Hebb}$. Furthermore, truncation saves computational resources and time, and more importantly, in experimental realizations, its validity witnesses a bound in the maximum excited state and thus in the operation energy (\cref{fig:memory}(c)). In any case, the experimental viability of our system mainly depends on the capacity to engineer an oscillator with a high nonlinear term. Superconducting resonators are a good candidate when $n = m$ due to their ability to realize any nonlinearity by modifying only the flux pump frequency \cite{lang2021multi,svensson2018period} with three-photon down-conversion achieved in \ccite{forn2020three}. Those systems have been used to generate catlike states by removing the linear dissipative term \cite{mundhada2017four}. Consequently, the appearance of the linear term makes it easier to realize in practice. Aside, experiments realizing phase-shifted coherent state discrimination have been pursued with success \cite{becerra2013experimental}.

To conclude, we believe this work heralds a new way of pursuing \gls{am} beyond typical spin chains. It would be interesting to see the robustness and scalability of this proposal when coupling a few nonlinear oscillators. More complex metastability scenarios where the spectral analysis is not sufficient could arise, e.g. in the presence of skin and topological effects \cite{haga2021liouvillian,mori2021metastability,flynn2021topology}. Also, in \ccite{cabot2021metastable}, it was shown that the onset of metastability relates to an exceptional point in the Liouvillian spectrum of the van der Pol oscillator. This and other dynamical aspects need to be further explored. An additional open question concerns the possibility of storing \textit{quantum memories}. While for the sake of clarity in this work we have focused on the case $n=m$, which is built around coherent-state discrimination, \cref{fig:lobes4}(c) shows that in different scenarios metastable squeezed states can emerge. This aspect is left for future work. 

\begin{acknowledgments}
We acknowledge the Spanish State Research Agency, through the Mar\'ia de Maeztu project CEX2021-001164-M funded by the MCIN/AEI/10.13039/501100011033 and through the QUARESC project (PID2019-109094GB-C21/AEI/ 10.13039/501100011033).
We also acknowledge funding by CAIB through the QUAREC project (PRD2018/47). The CSIC Interdisciplinary Thematic Platform (PTI) on Quantum Technologies in Spain is also acknowledged.
GLG is funded by the Spanish  Ministerio de Educaci\'on y Formaci\'on Profesional/Ministerio de Universidades and co-funded by the University of the Balearic Islands through the Beatriz Galindo program (BG20/00085). ALM is funded by the University of the Balearic Islands through the project BGRH-UIB-2021. We kindly acknowledge Albert Cabot for discussions and suggestions.
\end{acknowledgments}

\bibliography{references.bib}

\begin{thebibliography}{80}%
\makeatletter
\providecommand \@ifxundefined [1]{%
 \@ifx{#1\undefined}
}%
\providecommand \@ifnum [1]{%
 \ifnum #1\expandafter \@firstoftwo
 \else \expandafter \@secondoftwo
 \fi
}%
\providecommand \@ifx [1]{%
 \ifx #1\expandafter \@firstoftwo
 \else \expandafter \@secondoftwo
 \fi
}%
\providecommand \natexlab [1]{#1}%
\providecommand \enquote  [1]{``#1''}%
\providecommand \bibnamefont  [1]{#1}%
\providecommand \bibfnamefont [1]{#1}%
\providecommand \citenamefont [1]{#1}%
\providecommand \href@noop [0]{\@secondoftwo}%
\providecommand \href [0]{\begingroup \@sanitize@url \@href}%
\providecommand \@href[1]{\@@startlink{#1}\@@href}%
\providecommand \@@href[1]{\endgroup#1\@@endlink}%
\providecommand \@sanitize@url [0]{\catcode `\\12\catcode `\$12\catcode
  `\&12\catcode `\#12\catcode `\^12\catcode `\_12\catcode `\%12\relax}%
\providecommand \@@startlink[1]{}%
\providecommand \@@endlink[0]{}%
\providecommand \url  [0]{\begingroup\@sanitize@url \@url }%
\providecommand \@url [1]{\endgroup\@href {#1}{\urlprefix }}%
\providecommand \urlprefix  [0]{URL }%
\providecommand \Eprint [0]{\href }%
\providecommand \doibase [0]{https://doi.org/}%
\providecommand \selectlanguage [0]{\@gobble}%
\providecommand \bibinfo  [0]{\@secondoftwo}%
\providecommand \bibfield  [0]{\@secondoftwo}%
\providecommand \translation [1]{[#1]}%
\providecommand \BibitemOpen [0]{}%
\providecommand \bibitemStop [0]{}%
\providecommand \bibitemNoStop [0]{.\EOS\space}%
\providecommand \EOS [0]{\spacefactor3000\relax}%
\providecommand \BibitemShut  [1]{\csname bibitem#1\endcsname}%
\let\auto@bib@innerbib\@empty
%</preamble>
\bibitem [{\citenamefont {Amari}(1998)}]{amari1998gradient}%
  \BibitemOpen
  \bibfield  {author} {\bibinfo {author} {\bibfnamefont {S.-i.}\ \bibnamefont
  {Amari}},\ }\href {https://doi.org/10.1162/089976698300017746} {\bibfield
  {journal} {\bibinfo  {journal} {Neural Computation}\ }\textbf {\bibinfo
  {volume} {10}},\ \bibinfo {pages} {251} (\bibinfo {year} {1998})}\BibitemShut
  {NoStop}%
\bibitem [{\citenamefont {Bishop}\ and\ \citenamefont
  {Nasrabadi}(2006)}]{bishop2006pattern}%
  \BibitemOpen
  \bibfield  {author} {\bibinfo {author} {\bibfnamefont {C.~M.}\ \bibnamefont
  {Bishop}}\ and\ \bibinfo {author} {\bibfnamefont {N.~M.}\ \bibnamefont
  {Nasrabadi}},\ }\href@noop {} {\emph {\bibinfo {title} {Pattern recognition
  and machine learning}}},\ Vol.~\bibinfo {volume} {4}\ (\bibinfo  {publisher}
  {Springer},\ \bibinfo {year} {2006})\BibitemShut {NoStop}%
\bibitem [{\citenamefont {Hinton}\ and\ \citenamefont
  {Salakhutdinov}(2006)}]{hinton}%
  \BibitemOpen
  \bibfield  {author} {\bibinfo {author} {\bibfnamefont {G.~E.}\ \bibnamefont
  {Hinton}}\ and\ \bibinfo {author} {\bibfnamefont {R.~R.}\ \bibnamefont
  {Salakhutdinov}},\ }\href@noop {} {\bibfield  {journal} {\bibinfo  {journal}
  {Science}\ }\textbf {\bibinfo {volume} {313}},\ \bibinfo {pages} {504}
  (\bibinfo {year} {2006})}\BibitemShut {NoStop}%
\bibitem [{\citenamefont {Amit}(1989)}]{amit_1989}%
  \BibitemOpen
  \bibfield  {author} {\bibinfo {author} {\bibfnamefont {D.~J.}\ \bibnamefont
  {Amit}},\ }\href {https://doi.org/10.1017/CBO9780511623257} {\emph {\bibinfo
  {title} {Modeling Brain Function: The World of Attractor Neural Networks}}}\
  (\bibinfo  {publisher} {Cambridge University Press},\ \bibinfo {year}
  {1989})\BibitemShut {NoStop}%
\bibitem [{\citenamefont {Hertz}\ \emph {et~al.}(2018)\citenamefont {Hertz},
  \citenamefont {Krogh},\ and\ \citenamefont {Palmer}}]{hertz2018introduction}%
  \BibitemOpen
  \bibfield  {author} {\bibinfo {author} {\bibfnamefont {J.}~\bibnamefont
  {Hertz}}, \bibinfo {author} {\bibfnamefont {A.}~\bibnamefont {Krogh}},\ and\
  \bibinfo {author} {\bibfnamefont {R.~G.}\ \bibnamefont {Palmer}},\
  }\href@noop {} {\emph {\bibinfo {title} {Introduction to the theory of neural
  computation}}}\ (\bibinfo  {publisher} {CRC Press},\ \bibinfo {year}
  {2018})\BibitemShut {NoStop}%
\bibitem [{\citenamefont {Hopfield}(1982)}]{hopfield1982neural}%
  \BibitemOpen
  \bibfield  {author} {\bibinfo {author} {\bibfnamefont {J.~J.}\ \bibnamefont
  {Hopfield}},\ }\href@noop {} {\bibfield  {journal} {\bibinfo  {journal}
  {Proceedings of the national academy of sciences}\ }\textbf {\bibinfo
  {volume} {79}},\ \bibinfo {pages} {2554} (\bibinfo {year}
  {1982})}\BibitemShut {NoStop}%
\bibitem [{\citenamefont {Amit}\ \emph {et~al.}(1985)\citenamefont {Amit},
  \citenamefont {Gutfreund},\ and\ \citenamefont
  {Sompolinsky}}]{amit1985storing}%
  \BibitemOpen
  \bibfield  {author} {\bibinfo {author} {\bibfnamefont {D.~J.}\ \bibnamefont
  {Amit}}, \bibinfo {author} {\bibfnamefont {H.}~\bibnamefont {Gutfreund}},\
  and\ \bibinfo {author} {\bibfnamefont {H.}~\bibnamefont {Sompolinsky}},\
  }\href@noop {} {\bibfield  {journal} {\bibinfo  {journal} {Phys. Rev. Lett.}\
  }\textbf {\bibinfo {volume} {55}},\ \bibinfo {pages} {1530} (\bibinfo {year}
  {1985})}\BibitemShut {NoStop}%
\bibitem [{\citenamefont {Amit}\ \emph {et~al.}(1987)\citenamefont {Amit},
  \citenamefont {Gutfreund},\ and\ \citenamefont
  {Sompolinsky}}]{amit1987statistical}%
  \BibitemOpen
  \bibfield  {author} {\bibinfo {author} {\bibfnamefont {D.~J.}\ \bibnamefont
  {Amit}}, \bibinfo {author} {\bibfnamefont {H.}~\bibnamefont {Gutfreund}},\
  and\ \bibinfo {author} {\bibfnamefont {H.}~\bibnamefont {Sompolinsky}},\
  }\href@noop {} {\bibfield  {journal} {\bibinfo  {journal} {Annals of
  physics}\ }\textbf {\bibinfo {volume} {173}},\ \bibinfo {pages} {30}
  (\bibinfo {year} {1987})}\BibitemShut {NoStop}%
\bibitem [{\citenamefont {Wittek}(2014)}]{wittek2014quantum}%
  \BibitemOpen
  \bibfield  {author} {\bibinfo {author} {\bibfnamefont {P.}~\bibnamefont
  {Wittek}},\ }\href@noop {} {\emph {\bibinfo {title} {Quantum machine
  learning: what quantum computing means to data mining}}}\ (\bibinfo
  {publisher} {Academic Press},\ \bibinfo {year} {2014})\BibitemShut {NoStop}%
\bibitem [{\citenamefont {Biamonte}\ \emph {et~al.}(2017)\citenamefont
  {Biamonte}, \citenamefont {Wittek}, \citenamefont {Pancotti}, \citenamefont
  {Rebentrost}, \citenamefont {Wiebe},\ and\ \citenamefont
  {Lloyd}}]{biamonte2017quantum}%
  \BibitemOpen
  \bibfield  {author} {\bibinfo {author} {\bibfnamefont {J.}~\bibnamefont
  {Biamonte}}, \bibinfo {author} {\bibfnamefont {P.}~\bibnamefont {Wittek}},
  \bibinfo {author} {\bibfnamefont {N.}~\bibnamefont {Pancotti}}, \bibinfo
  {author} {\bibfnamefont {P.}~\bibnamefont {Rebentrost}}, \bibinfo {author}
  {\bibfnamefont {N.}~\bibnamefont {Wiebe}},\ and\ \bibinfo {author}
  {\bibfnamefont {S.}~\bibnamefont {Lloyd}},\ }\href@noop {} {\bibfield
  {journal} {\bibinfo  {journal} {Nature}\ }\textbf {\bibinfo {volume} {549}},\
  \bibinfo {pages} {195} (\bibinfo {year} {2017})}\BibitemShut {NoStop}%
\bibitem [{\citenamefont {Dunjko}\ and\ \citenamefont
  {Briegel}(2018)}]{Dunjko_2018}%
  \BibitemOpen
  \bibfield  {author} {\bibinfo {author} {\bibfnamefont {V.}~\bibnamefont
  {Dunjko}}\ and\ \bibinfo {author} {\bibfnamefont {H.~J.}\ \bibnamefont
  {Briegel}},\ }\href {https://doi.org/10.1088/1361-6633/aab406} {\bibfield
  {journal} {\bibinfo  {journal} {Rep. Prog. Phys.}\ }\textbf {\bibinfo
  {volume} {81}},\ \bibinfo {pages} {074001} (\bibinfo {year}
  {2018})}\BibitemShut {NoStop}%
\bibitem [{\citenamefont {Carleo}\ \emph {et~al.}(2019)\citenamefont {Carleo},
  \citenamefont {Cirac}, \citenamefont {Cranmer}, \citenamefont {Daudet},
  \citenamefont {Schuld}, \citenamefont {Tishby}, \citenamefont
  {Vogt-Maranto},\ and\ \citenamefont {Zdeborov\'a}}]{RevModPhys.91.045002}%
  \BibitemOpen
  \bibfield  {author} {\bibinfo {author} {\bibfnamefont {G.}~\bibnamefont
  {Carleo}}, \bibinfo {author} {\bibfnamefont {I.}~\bibnamefont {Cirac}},
  \bibinfo {author} {\bibfnamefont {K.}~\bibnamefont {Cranmer}}, \bibinfo
  {author} {\bibfnamefont {L.}~\bibnamefont {Daudet}}, \bibinfo {author}
  {\bibfnamefont {M.}~\bibnamefont {Schuld}}, \bibinfo {author} {\bibfnamefont
  {N.}~\bibnamefont {Tishby}}, \bibinfo {author} {\bibfnamefont
  {L.}~\bibnamefont {Vogt-Maranto}},\ and\ \bibinfo {author} {\bibfnamefont
  {L.}~\bibnamefont {Zdeborov\'a}},\ }\href
  {https://doi.org/10.1103/RevModPhys.91.045002} {\bibfield  {journal}
  {\bibinfo  {journal} {Rev. Mod. Phys.}\ }\textbf {\bibinfo {volume} {91}},\
  \bibinfo {pages} {045002} (\bibinfo {year} {2019})}\BibitemShut {NoStop}%
\bibitem [{\citenamefont {Gopalakrishnan}\ \emph {et~al.}(2012)\citenamefont
  {Gopalakrishnan}, \citenamefont {Lev},\ and\ \citenamefont
  {Goldbart}}]{Gopalakrishnan}%
  \BibitemOpen
  \bibfield  {author} {\bibinfo {author} {\bibfnamefont {S.}~\bibnamefont
  {Gopalakrishnan}}, \bibinfo {author} {\bibfnamefont {B.~L.}\ \bibnamefont
  {Lev}},\ and\ \bibinfo {author} {\bibfnamefont {P.~M.}\ \bibnamefont
  {Goldbart}},\ }\href {https://doi.org/10.1080/14786435.2011.637980}
  {\bibfield  {journal} {\bibinfo  {journal} {Philosophical Magazine}\ }\textbf
  {\bibinfo {volume} {92}},\ \bibinfo {pages} {353} (\bibinfo {year} {2012})},\
  \Eprint {https://arxiv.org/abs/https://doi.org/10.1080/14786435.2011.637980}
  {https://doi.org/10.1080/14786435.2011.637980} \BibitemShut {NoStop}%
\bibitem [{\citenamefont {Diamantini}\ and\ \citenamefont
  {Trugenberger}(2006)}]{diamantini2006quantum}%
  \BibitemOpen
  \bibfield  {author} {\bibinfo {author} {\bibfnamefont {M.~C.}\ \bibnamefont
  {Diamantini}}\ and\ \bibinfo {author} {\bibfnamefont {C.~A.}\ \bibnamefont
  {Trugenberger}},\ }\href@noop {} {\bibfield  {journal} {\bibinfo  {journal}
  {Physical review letters}\ }\textbf {\bibinfo {volume} {97}},\ \bibinfo
  {pages} {130503} (\bibinfo {year} {2006})}\BibitemShut {NoStop}%
\bibitem [{\citenamefont {Gopalakrishnan}\ \emph {et~al.}(2011)\citenamefont
  {Gopalakrishnan}, \citenamefont {Lev},\ and\ \citenamefont
  {Goldbart}}]{PhysRevLett.107.277201}%
  \BibitemOpen
  \bibfield  {author} {\bibinfo {author} {\bibfnamefont {S.}~\bibnamefont
  {Gopalakrishnan}}, \bibinfo {author} {\bibfnamefont {B.~L.}\ \bibnamefont
  {Lev}},\ and\ \bibinfo {author} {\bibfnamefont {P.~M.}\ \bibnamefont
  {Goldbart}},\ }\href {https://doi.org/10.1103/PhysRevLett.107.277201}
  {\bibfield  {journal} {\bibinfo  {journal} {Phys. Rev. Lett.}\ }\textbf
  {\bibinfo {volume} {107}},\ \bibinfo {pages} {277201} (\bibinfo {year}
  {2011})}\BibitemShut {NoStop}%
\bibitem [{\citenamefont {Torggler}\ \emph {et~al.}(2017)\citenamefont
  {Torggler}, \citenamefont {Kr\"amer},\ and\ \citenamefont
  {Ritsch}}]{PhysRevA.95.032310}%
  \BibitemOpen
  \bibfield  {author} {\bibinfo {author} {\bibfnamefont {V.}~\bibnamefont
  {Torggler}}, \bibinfo {author} {\bibfnamefont {S.}~\bibnamefont {Kr\"amer}},\
  and\ \bibinfo {author} {\bibfnamefont {H.}~\bibnamefont {Ritsch}},\ }\href
  {https://doi.org/10.1103/PhysRevA.95.032310} {\bibfield  {journal} {\bibinfo
  {journal} {Phys. Rev. A}\ }\textbf {\bibinfo {volume} {95}},\ \bibinfo
  {pages} {032310} (\bibinfo {year} {2017})}\BibitemShut {NoStop}%
\bibitem [{\citenamefont {Rotondo}\ \emph {et~al.}(2018)\citenamefont
  {Rotondo}, \citenamefont {Marcuzzi}, \citenamefont {Garrahan}, \citenamefont
  {Lesanovsky},\ and\ \citenamefont {M{\"u}ller}}]{rotondo2018open}%
  \BibitemOpen
  \bibfield  {author} {\bibinfo {author} {\bibfnamefont {P.}~\bibnamefont
  {Rotondo}}, \bibinfo {author} {\bibfnamefont {M.}~\bibnamefont {Marcuzzi}},
  \bibinfo {author} {\bibfnamefont {J.~P.}\ \bibnamefont {Garrahan}}, \bibinfo
  {author} {\bibfnamefont {I.}~\bibnamefont {Lesanovsky}},\ and\ \bibinfo
  {author} {\bibfnamefont {M.}~\bibnamefont {M{\"u}ller}},\ }\href@noop {}
  {\bibfield  {journal} {\bibinfo  {journal} {Journal of Physics A:
  Mathematical and Theoretical}\ }\textbf {\bibinfo {volume} {51}},\ \bibinfo
  {pages} {115301} (\bibinfo {year} {2018})}\BibitemShut {NoStop}%
\bibitem [{\citenamefont {Rebentrost}\ \emph {et~al.}(2018)\citenamefont
  {Rebentrost}, \citenamefont {Bromley}, \citenamefont {Weedbrook},\ and\
  \citenamefont {Lloyd}}]{PhysRevA.98.042308}%
  \BibitemOpen
  \bibfield  {author} {\bibinfo {author} {\bibfnamefont {P.}~\bibnamefont
  {Rebentrost}}, \bibinfo {author} {\bibfnamefont {T.~R.}\ \bibnamefont
  {Bromley}}, \bibinfo {author} {\bibfnamefont {C.}~\bibnamefont {Weedbrook}},\
  and\ \bibinfo {author} {\bibfnamefont {S.}~\bibnamefont {Lloyd}},\ }\href
  {https://doi.org/10.1103/PhysRevA.98.042308} {\bibfield  {journal} {\bibinfo
  {journal} {Phys. Rev. A}\ }\textbf {\bibinfo {volume} {98}},\ \bibinfo
  {pages} {042308} (\bibinfo {year} {2018})}\BibitemShut {NoStop}%
\bibitem [{\citenamefont {Carollo}\ and\ \citenamefont
  {Lesanovsky}(2021)}]{carollo2021exactness}%
  \BibitemOpen
  \bibfield  {author} {\bibinfo {author} {\bibfnamefont {F.}~\bibnamefont
  {Carollo}}\ and\ \bibinfo {author} {\bibfnamefont {I.}~\bibnamefont
  {Lesanovsky}},\ }\href@noop {} {\bibfield  {journal} {\bibinfo  {journal}
  {Physical Review Letters}\ }\textbf {\bibinfo {volume} {126}},\ \bibinfo
  {pages} {230601} (\bibinfo {year} {2021})}\BibitemShut {NoStop}%
\bibitem [{\citenamefont {Fiorelli}\ \emph {et~al.}(2022)\citenamefont
  {Fiorelli}, \citenamefont {Lesanovsky},\ and\ \citenamefont
  {M{\"u}ller}}]{fiorelli2021phase}%
  \BibitemOpen
  \bibfield  {author} {\bibinfo {author} {\bibfnamefont {E.}~\bibnamefont
  {Fiorelli}}, \bibinfo {author} {\bibfnamefont {I.}~\bibnamefont
  {Lesanovsky}},\ and\ \bibinfo {author} {\bibfnamefont {M.}~\bibnamefont
  {M{\"u}ller}},\ }\href@noop {} {\bibfield  {journal} {\bibinfo  {journal}
  {New Journal of Physics}\ } (\bibinfo {year} {2022})}\BibitemShut {NoStop}%
\bibitem [{\citenamefont {Marsh}\ \emph {et~al.}(2021)\citenamefont {Marsh},
  \citenamefont {Guo}, \citenamefont {Kroeze}, \citenamefont {Gopalakrishnan},
  \citenamefont {Ganguli}, \citenamefont {Keeling},\ and\ \citenamefont
  {Lev}}]{enhancing2021marsh}%
  \BibitemOpen
  \bibfield  {author} {\bibinfo {author} {\bibfnamefont {B.~P.}\ \bibnamefont
  {Marsh}}, \bibinfo {author} {\bibfnamefont {Y.}~\bibnamefont {Guo}}, \bibinfo
  {author} {\bibfnamefont {R.~M.}\ \bibnamefont {Kroeze}}, \bibinfo {author}
  {\bibfnamefont {S.}~\bibnamefont {Gopalakrishnan}}, \bibinfo {author}
  {\bibfnamefont {S.}~\bibnamefont {Ganguli}}, \bibinfo {author} {\bibfnamefont
  {J.}~\bibnamefont {Keeling}},\ and\ \bibinfo {author} {\bibfnamefont {B.~L.}\
  \bibnamefont {Lev}},\ }\href {https://doi.org/10.1103/PhysRevX.11.021048}
  {\bibfield  {journal} {\bibinfo  {journal} {Phys. Rev. X}\ }\textbf {\bibinfo
  {volume} {11}},\ \bibinfo {pages} {021048} (\bibinfo {year}
  {2021})}\BibitemShut {NoStop}%
\bibitem [{\citenamefont {Rotondo}\ \emph {et~al.}(2015)\citenamefont
  {Rotondo}, \citenamefont {Cosentino~Lagomarsino},\ and\ \citenamefont
  {Viola}}]{PhysRevLett.114.143601}%
  \BibitemOpen
  \bibfield  {author} {\bibinfo {author} {\bibfnamefont {P.}~\bibnamefont
  {Rotondo}}, \bibinfo {author} {\bibfnamefont {M.}~\bibnamefont
  {Cosentino~Lagomarsino}},\ and\ \bibinfo {author} {\bibfnamefont
  {G.}~\bibnamefont {Viola}},\ }\href
  {https://doi.org/10.1103/PhysRevLett.114.143601} {\bibfield  {journal}
  {\bibinfo  {journal} {Phys. Rev. Lett.}\ }\textbf {\bibinfo {volume} {114}},\
  \bibinfo {pages} {143601} (\bibinfo {year} {2015})}\BibitemShut {NoStop}%
\bibitem [{\citenamefont {Fiorelli}\ \emph {et~al.}(2020)\citenamefont
  {Fiorelli}, \citenamefont {Marcuzzi}, \citenamefont {Rotondo}, \citenamefont
  {Carollo},\ and\ \citenamefont {Lesanovsky}}]{fiorelli2020signatures}%
  \BibitemOpen
  \bibfield  {author} {\bibinfo {author} {\bibfnamefont {E.}~\bibnamefont
  {Fiorelli}}, \bibinfo {author} {\bibfnamefont {M.}~\bibnamefont {Marcuzzi}},
  \bibinfo {author} {\bibfnamefont {P.}~\bibnamefont {Rotondo}}, \bibinfo
  {author} {\bibfnamefont {F.}~\bibnamefont {Carollo}},\ and\ \bibinfo {author}
  {\bibfnamefont {I.}~\bibnamefont {Lesanovsky}},\ }\href@noop {} {\bibfield
  {journal} {\bibinfo  {journal} {Phys. Rev. Lett.}\ }\textbf {\bibinfo
  {volume} {125}},\ \bibinfo {pages} {070604} (\bibinfo {year}
  {2020})}\BibitemShut {NoStop}%
\bibitem [{\citenamefont {Lewenstein}\ \emph {et~al.}(2021)\citenamefont
  {Lewenstein}, \citenamefont {Gratsea}, \citenamefont {Riera-Campeny},
  \citenamefont {Aloy}, \citenamefont {Kasper},\ and\ \citenamefont
  {Sanpera}}]{sanpera2021capacity}%
  \BibitemOpen
  \bibfield  {author} {\bibinfo {author} {\bibfnamefont {M.}~\bibnamefont
  {Lewenstein}}, \bibinfo {author} {\bibfnamefont {A.}~\bibnamefont {Gratsea}},
  \bibinfo {author} {\bibfnamefont {A.}~\bibnamefont {Riera-Campeny}}, \bibinfo
  {author} {\bibfnamefont {A.}~\bibnamefont {Aloy}}, \bibinfo {author}
  {\bibfnamefont {V.}~\bibnamefont {Kasper}},\ and\ \bibinfo {author}
  {\bibfnamefont {A.}~\bibnamefont {Sanpera}},\ }\href
  {https://doi.org/10.1088/2058-9565/ac070f} {\bibfield  {journal} {\bibinfo
  {journal} {Quantum Science and Technology}\ }\textbf {\bibinfo {volume}
  {6}},\ \bibinfo {pages} {045002} (\bibinfo {year} {2021})}\BibitemShut
  {NoStop}%
\bibitem [{\citenamefont {B{\"o}deker}\ \emph {et~al.}(2022)\citenamefont
  {B{\"o}deker}, \citenamefont {Fiorelli},\ and\ \citenamefont
  {M{\"u}ller}}]{bodeker2022optimal}%
  \BibitemOpen
  \bibfield  {author} {\bibinfo {author} {\bibfnamefont {L.}~\bibnamefont
  {B{\"o}deker}}, \bibinfo {author} {\bibfnamefont {E.}~\bibnamefont
  {Fiorelli}},\ and\ \bibinfo {author} {\bibfnamefont {M.}~\bibnamefont
  {M{\"u}ller}},\ }\href@noop {} {\bibfield  {journal} {\bibinfo  {journal}
  {arXiv preprint arXiv:2210.07894}\ } (\bibinfo {year} {2022})}\BibitemShut
  {NoStop}%
\bibitem [{\citenamefont {Gardner}(1988)}]{gardner1988space}%
  \BibitemOpen
  \bibfield  {author} {\bibinfo {author} {\bibfnamefont {E.}~\bibnamefont
  {Gardner}},\ }\href@noop {} {\bibfield  {journal} {\bibinfo  {journal}
  {Journal of physics A: Mathematical and general}\ }\textbf {\bibinfo {volume}
  {21}},\ \bibinfo {pages} {257} (\bibinfo {year} {1988})}\BibitemShut
  {NoStop}%
\bibitem [{\citenamefont {Gardner}\ and\ \citenamefont
  {Derrida}(1988)}]{gardner1988optimal}%
  \BibitemOpen
  \bibfield  {author} {\bibinfo {author} {\bibfnamefont {E.}~\bibnamefont
  {Gardner}}\ and\ \bibinfo {author} {\bibfnamefont {B.}~\bibnamefont
  {Derrida}},\ }\href@noop {} {\bibfield  {journal} {\bibinfo  {journal}
  {Journal of Physics A: Mathematical and general}\ }\textbf {\bibinfo {volume}
  {21}},\ \bibinfo {pages} {271} (\bibinfo {year} {1988})}\BibitemShut
  {NoStop}%
\bibitem [{\citenamefont {Gratsea}\ \emph {et~al.}(2021)\citenamefont
  {Gratsea}, \citenamefont {Kasper},\ and\ \citenamefont
  {Lewenstein}}]{gratsea2021storage}%
  \BibitemOpen
  \bibfield  {author} {\bibinfo {author} {\bibfnamefont {A.}~\bibnamefont
  {Gratsea}}, \bibinfo {author} {\bibfnamefont {V.}~\bibnamefont {Kasper}},\
  and\ \bibinfo {author} {\bibfnamefont {M.}~\bibnamefont {Lewenstein}},\
  }\href@noop {} {\bibfield  {journal} {\bibinfo  {journal} {arXiv preprint
  arXiv:2111.08414}\ } (\bibinfo {year} {2021})}\BibitemShut {NoStop}%
\bibitem [{\citenamefont {Benatti}\ \emph {et~al.}(2022)\citenamefont
  {Benatti}, \citenamefont {Gramegna},\ and\ \citenamefont
  {Mancini}}]{benatti2022pattern}%
  \BibitemOpen
  \bibfield  {author} {\bibinfo {author} {\bibfnamefont {F.}~\bibnamefont
  {Benatti}}, \bibinfo {author} {\bibfnamefont {G.}~\bibnamefont {Gramegna}},\
  and\ \bibinfo {author} {\bibfnamefont {S.}~\bibnamefont {Mancini}},\
  }\href@noop {} {\bibfield  {journal} {\bibinfo  {journal} {Journal of Physics
  A: Mathematical and Theoretical}\ } (\bibinfo {year} {2022})}\BibitemShut
  {NoStop}%
\bibitem [{\citenamefont {Brinkman}\ \emph {et~al.}(2022)\citenamefont
  {Brinkman}, \citenamefont {Yan}, \citenamefont {Maffei}, \citenamefont
  {Park}, \citenamefont {Fontanini}, \citenamefont {Wang},\ and\ \citenamefont
  {La~Camera}}]{brinkman2022metastable}%
  \BibitemOpen
  \bibfield  {author} {\bibinfo {author} {\bibfnamefont {B.~A.}\ \bibnamefont
  {Brinkman}}, \bibinfo {author} {\bibfnamefont {H.}~\bibnamefont {Yan}},
  \bibinfo {author} {\bibfnamefont {A.}~\bibnamefont {Maffei}}, \bibinfo
  {author} {\bibfnamefont {I.~M.}\ \bibnamefont {Park}}, \bibinfo {author}
  {\bibfnamefont {A.}~\bibnamefont {Fontanini}}, \bibinfo {author}
  {\bibfnamefont {J.}~\bibnamefont {Wang}},\ and\ \bibinfo {author}
  {\bibfnamefont {G.}~\bibnamefont {La~Camera}},\ }\href@noop {} {\bibfield
  {journal} {\bibinfo  {journal} {Applied Physics Reviews}\ }\textbf {\bibinfo
  {volume} {9}},\ \bibinfo {pages} {011313} (\bibinfo {year}
  {2022})}\BibitemShut {NoStop}%
\bibitem [{\citenamefont {Govia}\ \emph {et~al.}(2021)\citenamefont {Govia},
  \citenamefont {Ribeill}, \citenamefont {Rowlands}, \citenamefont {Krovi},\
  and\ \citenamefont {Ohki}}]{govia2021quantum}%
  \BibitemOpen
  \bibfield  {author} {\bibinfo {author} {\bibfnamefont {L.~C.~G.}\
  \bibnamefont {Govia}}, \bibinfo {author} {\bibfnamefont {G.~J.}\ \bibnamefont
  {Ribeill}}, \bibinfo {author} {\bibfnamefont {G.~E.}\ \bibnamefont
  {Rowlands}}, \bibinfo {author} {\bibfnamefont {H.~K.}\ \bibnamefont
  {Krovi}},\ and\ \bibinfo {author} {\bibfnamefont {T.~A.}\ \bibnamefont
  {Ohki}},\ }\href {https://doi.org/10.1103/PhysRevResearch.3.013077}
  {\bibfield  {journal} {\bibinfo  {journal} {Phys. Rev. Research}\ }\textbf
  {\bibinfo {volume} {3}},\ \bibinfo {pages} {013077} (\bibinfo {year}
  {2021})}\BibitemShut {NoStop}%
\bibitem [{\citenamefont {Minganti}\ \emph {et~al.}(2018)\citenamefont
  {Minganti}, \citenamefont {Biella}, \citenamefont {Bartolo},\ and\
  \citenamefont {Ciuti}}]{minganti2018spectral}%
  \BibitemOpen
  \bibfield  {author} {\bibinfo {author} {\bibfnamefont {F.}~\bibnamefont
  {Minganti}}, \bibinfo {author} {\bibfnamefont {A.}~\bibnamefont {Biella}},
  \bibinfo {author} {\bibfnamefont {N.}~\bibnamefont {Bartolo}},\ and\ \bibinfo
  {author} {\bibfnamefont {C.}~\bibnamefont {Ciuti}},\ }\href@noop {}
  {\bibfield  {journal} {\bibinfo  {journal} {Phys. Rev. A}\ }\textbf {\bibinfo
  {volume} {98}},\ \bibinfo {pages} {042118} (\bibinfo {year}
  {2018})}\BibitemShut {NoStop}%
\bibitem [{\citenamefont {Macieszczak}\ \emph {et~al.}(2016)\citenamefont
  {Macieszczak}, \citenamefont {Gu{\c{t}}{\u{a}}}, \citenamefont {Lesanovsky},\
  and\ \citenamefont {Garrahan}}]{macieszczak2016towards}%
  \BibitemOpen
  \bibfield  {author} {\bibinfo {author} {\bibfnamefont {K.}~\bibnamefont
  {Macieszczak}}, \bibinfo {author} {\bibfnamefont {M.}~\bibnamefont
  {Gu{\c{t}}{\u{a}}}}, \bibinfo {author} {\bibfnamefont {I.}~\bibnamefont
  {Lesanovsky}},\ and\ \bibinfo {author} {\bibfnamefont {J.~P.}\ \bibnamefont
  {Garrahan}},\ }\href@noop {} {\bibfield  {journal} {\bibinfo  {journal}
  {Phys. Rev. Lett.}\ }\textbf {\bibinfo {volume} {116}},\ \bibinfo {pages}
  {240404} (\bibinfo {year} {2016})}\BibitemShut {NoStop}%
\bibitem [{\citenamefont {Cabot}\ \emph {et~al.}(2021)\citenamefont {Cabot},
  \citenamefont {Giorgi},\ and\ \citenamefont
  {Zambrini}}]{cabot2021metastable}%
  \BibitemOpen
  \bibfield  {author} {\bibinfo {author} {\bibfnamefont {A.}~\bibnamefont
  {Cabot}}, \bibinfo {author} {\bibfnamefont {G.~L.}\ \bibnamefont {Giorgi}},\
  and\ \bibinfo {author} {\bibfnamefont {R.}~\bibnamefont {Zambrini}},\
  }\href@noop {} {\bibfield  {journal} {\bibinfo  {journal} {New Journal of
  Physics}\ } (\bibinfo {year} {2021})}\BibitemShut {NoStop}%
\bibitem [{\citenamefont {Lindblad}(1976)}]{lindblad1976generators}%
  \BibitemOpen
  \bibfield  {author} {\bibinfo {author} {\bibfnamefont {G.}~\bibnamefont
  {Lindblad}},\ }\href@noop {} {\bibfield  {journal} {\bibinfo  {journal}
  {Communications in Mathematical Physics}\ }\textbf {\bibinfo {volume} {48}},\
  \bibinfo {pages} {119} (\bibinfo {year} {1976})}\BibitemShut {NoStop}%
\bibitem [{\citenamefont {Gorini}\ \emph {et~al.}(1976)\citenamefont {Gorini},
  \citenamefont {Kossakowski},\ and\ \citenamefont
  {Sudarshan}}]{gorini1976completely}%
  \BibitemOpen
  \bibfield  {author} {\bibinfo {author} {\bibfnamefont {V.}~\bibnamefont
  {Gorini}}, \bibinfo {author} {\bibfnamefont {A.}~\bibnamefont
  {Kossakowski}},\ and\ \bibinfo {author} {\bibfnamefont {E.~C.~G.}\
  \bibnamefont {Sudarshan}},\ }\href@noop {} {\bibfield  {journal} {\bibinfo
  {journal} {Journal of Mathematical Physics}\ }\textbf {\bibinfo {volume}
  {17}},\ \bibinfo {pages} {821} (\bibinfo {year} {1976})}\BibitemShut
  {NoStop}%
\bibitem [{\citenamefont {Evans}\ and\ \citenamefont
  {Hanche-Olsen}(1979)}]{evans1977generators}%
  \BibitemOpen
  \bibfield  {author} {\bibinfo {author} {\bibfnamefont {D.~E.}\ \bibnamefont
  {Evans}}\ and\ \bibinfo {author} {\bibfnamefont {H.}~\bibnamefont
  {Hanche-Olsen}},\ }\href
  {https://doi.org/https://doi.org/10.1016/0022-1236(79)90054-5} {\bibfield
  {journal} {\bibinfo  {journal} {Journal of Functional Analysis}\ }\textbf
  {\bibinfo {volume} {32}},\ \bibinfo {pages} {207} (\bibinfo {year}
  {1979})}\BibitemShut {NoStop}%
\bibitem [{\citenamefont {Baumgartner}\ and\ \citenamefont
  {Narnhofer}(2008)}]{baumgartner2008analysis}%
  \BibitemOpen
  \bibfield  {author} {\bibinfo {author} {\bibfnamefont {B.}~\bibnamefont
  {Baumgartner}}\ and\ \bibinfo {author} {\bibfnamefont {H.}~\bibnamefont
  {Narnhofer}},\ }\href@noop {} {\bibfield  {journal} {\bibinfo  {journal}
  {Journal of Physics A: Mathematical and Theoretical}\ }\textbf {\bibinfo
  {volume} {41}},\ \bibinfo {pages} {395303} (\bibinfo {year}
  {2008})}\BibitemShut {NoStop}%
\bibitem [{\citenamefont {Macieszczak}\ \emph {et~al.}(2021)\citenamefont
  {Macieszczak}, \citenamefont {Rose}, \citenamefont {Lesanovsky},\ and\
  \citenamefont {Garrahan}}]{macieszczak2021theory}%
  \BibitemOpen
  \bibfield  {author} {\bibinfo {author} {\bibfnamefont {K.}~\bibnamefont
  {Macieszczak}}, \bibinfo {author} {\bibfnamefont {D.~C.}\ \bibnamefont
  {Rose}}, \bibinfo {author} {\bibfnamefont {I.}~\bibnamefont {Lesanovsky}},\
  and\ \bibinfo {author} {\bibfnamefont {J.~P.}\ \bibnamefont {Garrahan}},\
  }\href@noop {} {\bibfield  {journal} {\bibinfo  {journal} {Phys. Rev. Res.}\
  }\textbf {\bibinfo {volume} {3}},\ \bibinfo {pages} {033047} (\bibinfo {year}
  {2021})}\BibitemShut {NoStop}%
\bibitem [{Sup()}]{SuppMaterial}%
  \BibitemOpen
  \href@noop {} {\bibinfo {title} {See supplemental material for
  details on analytical calculations, which includes refs.
  \cite{rudolph2003unambiguous,azouit2015convergence,azouit2016well,chamberland2022building,meccia2002biologically,carmichael1999statistical}.}}\BibitemShut
  {Stop}%
\bibitem [{\citenamefont {Mundhada}\ \emph {et~al.}(2017)\citenamefont
  {Mundhada}, \citenamefont {Grimm}, \citenamefont {Touzard}, \citenamefont
  {Vool}, \citenamefont {Shankar}, \citenamefont {Devoret},\ and\ \citenamefont
  {Mirrahimi}}]{mundhada2017four}%
  \BibitemOpen
  \bibfield  {author} {\bibinfo {author} {\bibfnamefont {S.~O.}\ \bibnamefont
  {Mundhada}}, \bibinfo {author} {\bibfnamefont {A.}~\bibnamefont {Grimm}},
  \bibinfo {author} {\bibfnamefont {S.}~\bibnamefont {Touzard}}, \bibinfo
  {author} {\bibfnamefont {U.}~\bibnamefont {Vool}}, \bibinfo {author}
  {\bibfnamefont {S.}~\bibnamefont {Shankar}}, \bibinfo {author} {\bibfnamefont
  {M.~H.}\ \bibnamefont {Devoret}},\ and\ \bibinfo {author} {\bibfnamefont
  {M.}~\bibnamefont {Mirrahimi}},\ }\href
  {https://doi.org/10.1088/2058-9565/aa6e9d} {\bibfield  {journal} {\bibinfo
  {journal} {Quantum Science and Technology}\ }\textbf {\bibinfo {volume}
  {2}},\ \bibinfo {pages} {024005} (\bibinfo {year} {2017})}\BibitemShut
  {NoStop}%
\bibitem [{\citenamefont {Gevorkyan}\ and\ \citenamefont
  {Chaltykyan}(1999)}]{gevorkyan1999coherent}%
  \BibitemOpen
  \bibfield  {author} {\bibinfo {author} {\bibfnamefont {S.}~\bibnamefont
  {Gevorkyan}}\ and\ \bibinfo {author} {\bibfnamefont {V.}~\bibnamefont
  {Chaltykyan}},\ }\href@noop {} {\bibfield  {journal} {\bibinfo  {journal}
  {Journal of Modern Optics}\ }\textbf {\bibinfo {volume} {46}},\ \bibinfo
  {pages} {1447} (\bibinfo {year} {1999})}\BibitemShut {NoStop}%
\bibitem [{\citenamefont {Braunstein}\ and\ \citenamefont
  {McLachlan}(1987)}]{braunstein1987generalized}%
  \BibitemOpen
  \bibfield  {author} {\bibinfo {author} {\bibfnamefont {S.~L.}\ \bibnamefont
  {Braunstein}}\ and\ \bibinfo {author} {\bibfnamefont {R.~I.}\ \bibnamefont
  {McLachlan}},\ }\href@noop {} {\bibfield  {journal} {\bibinfo  {journal}
  {Physical Review A}\ }\textbf {\bibinfo {volume} {35}},\ \bibinfo {pages}
  {1659} (\bibinfo {year} {1987})}\BibitemShut {NoStop}%
\bibitem [{\citenamefont {Lang}\ and\ \citenamefont
  {Armour}(2021)}]{lang2021multi}%
  \BibitemOpen
  \bibfield  {author} {\bibinfo {author} {\bibfnamefont {B.}~\bibnamefont
  {Lang}}\ and\ \bibinfo {author} {\bibfnamefont {A.~D.}\ \bibnamefont
  {Armour}},\ }\href@noop {} {\bibfield  {journal} {\bibinfo  {journal} {New
  Journal of Physics}\ }\textbf {\bibinfo {volume} {23}},\ \bibinfo {pages}
  {033021} (\bibinfo {year} {2021})}\BibitemShut {NoStop}%
\bibitem [{\citenamefont {Minganti}\ \emph {et~al.}(2023)\citenamefont
  {Minganti}, \citenamefont {Savona},\ and\ \citenamefont
  {Biella}}]{minganti2023dissipative}%
  \BibitemOpen
  \bibfield  {author} {\bibinfo {author} {\bibfnamefont {F.}~\bibnamefont
  {Minganti}}, \bibinfo {author} {\bibfnamefont {V.}~\bibnamefont {Savona}},\
  and\ \bibinfo {author} {\bibfnamefont {A.}~\bibnamefont {Biella}},\
  }\href@noop {} {\bibinfo {title} {Dissipative phase transitions in $n$-photon
  driven quantum nonlinear resonators}} (\bibinfo {year} {2023}),\ \Eprint
  {https://arxiv.org/abs/2303.03355} {arXiv:2303.03355 [quant-ph]} \BibitemShut
  {NoStop}%
\bibitem [{\citenamefont {Bartolo}\ \emph {et~al.}(2016)\citenamefont
  {Bartolo}, \citenamefont {Minganti}, \citenamefont {Casteels},\ and\
  \citenamefont {Ciuti}}]{bartolo2016exact}%
  \BibitemOpen
  \bibfield  {author} {\bibinfo {author} {\bibfnamefont {N.}~\bibnamefont
  {Bartolo}}, \bibinfo {author} {\bibfnamefont {F.}~\bibnamefont {Minganti}},
  \bibinfo {author} {\bibfnamefont {W.}~\bibnamefont {Casteels}},\ and\
  \bibinfo {author} {\bibfnamefont {C.}~\bibnamefont {Ciuti}},\ }\href
  {https://doi.org/10.1103/PhysRevA.94.033841} {\bibfield  {journal} {\bibinfo
  {journal} {Phys. Rev. A}\ }\textbf {\bibinfo {volume} {94}},\ \bibinfo
  {pages} {033841} (\bibinfo {year} {2016})}\BibitemShut {NoStop}%
\bibitem [{\citenamefont {Mirrahimi}\ \emph {et~al.}(2014)\citenamefont
  {Mirrahimi}, \citenamefont {Leghtas}, \citenamefont {Albert}, \citenamefont
  {Touzard}, \citenamefont {Schoelkopf}, \citenamefont {Jiang},\ and\
  \citenamefont {Devoret}}]{mirrahimi2014dynamically}%
  \BibitemOpen
  \bibfield  {author} {\bibinfo {author} {\bibfnamefont {M.}~\bibnamefont
  {Mirrahimi}}, \bibinfo {author} {\bibfnamefont {Z.}~\bibnamefont {Leghtas}},
  \bibinfo {author} {\bibfnamefont {V.~V.}\ \bibnamefont {Albert}}, \bibinfo
  {author} {\bibfnamefont {S.}~\bibnamefont {Touzard}}, \bibinfo {author}
  {\bibfnamefont {R.~J.}\ \bibnamefont {Schoelkopf}}, \bibinfo {author}
  {\bibfnamefont {L.}~\bibnamefont {Jiang}},\ and\ \bibinfo {author}
  {\bibfnamefont {M.~H.}\ \bibnamefont {Devoret}},\ }\href@noop {} {\bibfield
  {journal} {\bibinfo  {journal} {New Journal of Physics}\ }\textbf {\bibinfo
  {volume} {16}},\ \bibinfo {pages} {045014} (\bibinfo {year}
  {2014})}\BibitemShut {NoStop}%
\bibitem [{\citenamefont {Ma}\ \emph {et~al.}(2021)\citenamefont {Ma},
  \citenamefont {Puri}, \citenamefont {Schoelkopf}, \citenamefont {Devoret},
  \citenamefont {Girvin},\ and\ \citenamefont {Jiang}}]{ma2021quantum}%
  \BibitemOpen
  \bibfield  {author} {\bibinfo {author} {\bibfnamefont {W.-L.}\ \bibnamefont
  {Ma}}, \bibinfo {author} {\bibfnamefont {S.}~\bibnamefont {Puri}}, \bibinfo
  {author} {\bibfnamefont {R.~J.}\ \bibnamefont {Schoelkopf}}, \bibinfo
  {author} {\bibfnamefont {M.~H.}\ \bibnamefont {Devoret}}, \bibinfo {author}
  {\bibfnamefont {S.}~\bibnamefont {Girvin}},\ and\ \bibinfo {author}
  {\bibfnamefont {L.}~\bibnamefont {Jiang}},\ }\href@noop {} {\bibfield
  {journal} {\bibinfo  {journal} {Science Bulletin}\ }\textbf {\bibinfo
  {volume} {66}},\ \bibinfo {pages} {1789} (\bibinfo {year}
  {2021})}\BibitemShut {NoStop}%
\bibitem [{\citenamefont {Gevorgyan}\ \emph {et~al.}(2008)\citenamefont
  {Gevorgyan}, \citenamefont {Xiao},\ and\ \citenamefont
  {Chaltykyan}}]{gevorgyan2008superposition}%
  \BibitemOpen
  \bibfield  {author} {\bibinfo {author} {\bibfnamefont {S.}~\bibnamefont
  {Gevorgyan}}, \bibinfo {author} {\bibfnamefont {M.}~\bibnamefont {Xiao}},\
  and\ \bibinfo {author} {\bibfnamefont {V.}~\bibnamefont {Chaltykyan}},\
  }\href@noop {} {\bibfield  {journal} {\bibinfo  {journal} {Journal of Modern
  Optics}\ }\textbf {\bibinfo {volume} {55}},\ \bibinfo {pages} {1923}
  (\bibinfo {year} {2008})}\BibitemShut {NoStop}%
\bibitem [{num()}]{numerics}%
  \BibitemOpen
  \href@noop {} {\bibinfo {title} {All the results presented have been obtained
  numerically using \texttt{QuTip} \cite{Johansson2013qutip} and
  \texttt{QuantumOptics.jl} \cite{kramer2018quantumoptics}, code available at
  \href{https://gitlab.ifisc.uib-csic.es/quantum/quantum-associative-memory-with-a-single-driven-dissipative-non-linear-oscillator}{https://gitlab.ifisc.uib-csic.es/quantum/}.}}\BibitemShut
  {Stop}%
\bibitem [{\citenamefont {Gilles}\ \emph {et~al.}(1994)\citenamefont {Gilles},
  \citenamefont {Garraway},\ and\ \citenamefont
  {Knight}}]{gilles1994generation}%
  \BibitemOpen
  \bibfield  {author} {\bibinfo {author} {\bibfnamefont {L.}~\bibnamefont
  {Gilles}}, \bibinfo {author} {\bibfnamefont {B.~M.}\ \bibnamefont
  {Garraway}},\ and\ \bibinfo {author} {\bibfnamefont {P.~L.}\ \bibnamefont
  {Knight}},\ }\href {https://doi.org/10.1103/PhysRevA.49.2785} {\bibfield
  {journal} {\bibinfo  {journal} {Phys. Rev. A}\ }\textbf {\bibinfo {volume}
  {49}},\ \bibinfo {pages} {2785} (\bibinfo {year} {1994})}\BibitemShut
  {NoStop}%
\bibitem [{Note1()}]{Note1}%
  \BibitemOpen
  \bibinfo {note} {This is in correspondence with the results found in
  Ref.~\cite {sonar2018squeezing} where squeezing reduces the effects of noise
  and enhances the stability of the lobes.}\BibitemShut {Stop}%
\bibitem [{\citenamefont {Mok}\ \emph {et~al.}(2020)\citenamefont {Mok},
  \citenamefont {Kwek},\ and\ \citenamefont
  {Heimonen}}]{PhysRevResearch.2.033422}%
  \BibitemOpen
  \bibfield  {author} {\bibinfo {author} {\bibfnamefont {W.-K.}\ \bibnamefont
  {Mok}}, \bibinfo {author} {\bibfnamefont {L.-C.}\ \bibnamefont {Kwek}},\ and\
  \bibinfo {author} {\bibfnamefont {H.}~\bibnamefont {Heimonen}},\ }\href
  {https://doi.org/10.1103/PhysRevResearch.2.033422} {\bibfield  {journal}
  {\bibinfo  {journal} {Phys. Rev. Res.}\ }\textbf {\bibinfo {volume} {2}},\
  \bibinfo {pages} {033422} (\bibinfo {year} {2020})}\BibitemShut {NoStop}%
\bibitem [{\citenamefont {Svensson}\ \emph {et~al.}(2018)\citenamefont
  {Svensson}, \citenamefont {Bengtsson}, \citenamefont {Bylander},
  \citenamefont {Shumeiko},\ and\ \citenamefont
  {Delsing}}]{svensson2018period}%
  \BibitemOpen
  \bibfield  {author} {\bibinfo {author} {\bibfnamefont {I.-M.}\ \bibnamefont
  {Svensson}}, \bibinfo {author} {\bibfnamefont {A.}~\bibnamefont {Bengtsson}},
  \bibinfo {author} {\bibfnamefont {J.}~\bibnamefont {Bylander}}, \bibinfo
  {author} {\bibfnamefont {V.}~\bibnamefont {Shumeiko}},\ and\ \bibinfo
  {author} {\bibfnamefont {P.}~\bibnamefont {Delsing}},\ }\href@noop {}
  {\bibfield  {journal} {\bibinfo  {journal} {App. Phys. Lett.}\ }\textbf
  {\bibinfo {volume} {113}},\ \bibinfo {pages} {022602} (\bibinfo {year}
  {2018})}\BibitemShut {NoStop}%
\bibitem [{\citenamefont {Sonar}\ \emph {et~al.}(2018)\citenamefont {Sonar},
  \citenamefont {Hajdu{\v{s}}ek}, \citenamefont {Mukherjee}, \citenamefont
  {Fazio}, \citenamefont {Vedral}, \citenamefont {Vinjanampathy},\ and\
  \citenamefont {Kwek}}]{sonar2018squeezing}%
  \BibitemOpen
  \bibfield  {author} {\bibinfo {author} {\bibfnamefont {S.}~\bibnamefont
  {Sonar}}, \bibinfo {author} {\bibfnamefont {M.}~\bibnamefont
  {Hajdu{\v{s}}ek}}, \bibinfo {author} {\bibfnamefont {M.}~\bibnamefont
  {Mukherjee}}, \bibinfo {author} {\bibfnamefont {R.}~\bibnamefont {Fazio}},
  \bibinfo {author} {\bibfnamefont {V.}~\bibnamefont {Vedral}}, \bibinfo
  {author} {\bibfnamefont {S.}~\bibnamefont {Vinjanampathy}},\ and\ \bibinfo
  {author} {\bibfnamefont {L.-C.}\ \bibnamefont {Kwek}},\ }\href@noop {}
  {\bibfield  {journal} {\bibinfo  {journal} {Phys. Rev. Lett.}\ }\textbf
  {\bibinfo {volume} {120}},\ \bibinfo {pages} {163601} (\bibinfo {year}
  {2018})}\BibitemShut {NoStop}%
\bibitem [{\citenamefont {Nair}\ \emph {et~al.}(2012)\citenamefont {Nair},
  \citenamefont {Yen}, \citenamefont {Guha}, \citenamefont {Shapiro},\ and\
  \citenamefont {Pirandola}}]{nair_symmetric_2012}%
  \BibitemOpen
  \bibfield  {author} {\bibinfo {author} {\bibfnamefont {R.}~\bibnamefont
  {Nair}}, \bibinfo {author} {\bibfnamefont {B.~J.}\ \bibnamefont {Yen}},
  \bibinfo {author} {\bibfnamefont {S.}~\bibnamefont {Guha}}, \bibinfo {author}
  {\bibfnamefont {J.~H.}\ \bibnamefont {Shapiro}},\ and\ \bibinfo {author}
  {\bibfnamefont {S.}~\bibnamefont {Pirandola}},\ }\href
  {https://doi.org/10.1103/PhysRevA.86.022306} {\bibfield  {journal} {\bibinfo
  {journal} {Phys. Rev. A}\ }\textbf {\bibinfo {volume} {86}},\ \bibinfo
  {pages} {022306} (\bibinfo {year} {2012})}\BibitemShut {NoStop}%
\bibitem [{\citenamefont {Izumi}\ \emph {et~al.}(2013)\citenamefont {Izumi},
  \citenamefont {Takeoka}, \citenamefont {Ema},\ and\ \citenamefont
  {Sasaki}}]{izumi_quantum_2013}%
  \BibitemOpen
  \bibfield  {author} {\bibinfo {author} {\bibfnamefont {S.}~\bibnamefont
  {Izumi}}, \bibinfo {author} {\bibfnamefont {M.}~\bibnamefont {Takeoka}},
  \bibinfo {author} {\bibfnamefont {K.}~\bibnamefont {Ema}},\ and\ \bibinfo
  {author} {\bibfnamefont {M.}~\bibnamefont {Sasaki}},\ }\href
  {https://doi.org/10.1103/PhysRevA.87.042328} {\bibfield  {journal} {\bibinfo
  {journal} {Phys. Rev. A}\ }\textbf {\bibinfo {volume} {87}},\ \bibinfo
  {pages} {042328} (\bibinfo {year} {2013})}\BibitemShut {NoStop}%
\bibitem [{\citenamefont {L\"orch}\ \emph {et~al.}(2019)\citenamefont
  {L\"orch}, \citenamefont {Zhang}, \citenamefont {Bruder},\ and\ \citenamefont
  {Dykman}}]{lorch2019quantum}%
  \BibitemOpen
  \bibfield  {author} {\bibinfo {author} {\bibfnamefont {N.}~\bibnamefont
  {L\"orch}}, \bibinfo {author} {\bibfnamefont {Y.}~\bibnamefont {Zhang}},
  \bibinfo {author} {\bibfnamefont {C.}~\bibnamefont {Bruder}},\ and\ \bibinfo
  {author} {\bibfnamefont {M.~I.}\ \bibnamefont {Dykman}},\ }\href
  {https://doi.org/10.1103/PhysRevResearch.1.023023} {\bibfield  {journal}
  {\bibinfo  {journal} {Phys. Rev. Res.}\ }\textbf {\bibinfo {volume} {1}},\
  \bibinfo {pages} {023023} (\bibinfo {year} {2019})}\BibitemShut {NoStop}%
\bibitem [{\citenamefont {Izumi}\ \emph {et~al.}(2012)\citenamefont {Izumi},
  \citenamefont {Takeoka}, \citenamefont {Fujiwara}, \citenamefont {Pozza},
  \citenamefont {Assalini}, \citenamefont {Ema},\ and\ \citenamefont
  {Sasaki}}]{izumi2012displacement}%
  \BibitemOpen
  \bibfield  {author} {\bibinfo {author} {\bibfnamefont {S.}~\bibnamefont
  {Izumi}}, \bibinfo {author} {\bibfnamefont {M.}~\bibnamefont {Takeoka}},
  \bibinfo {author} {\bibfnamefont {M.}~\bibnamefont {Fujiwara}}, \bibinfo
  {author} {\bibfnamefont {N.~D.}\ \bibnamefont {Pozza}}, \bibinfo {author}
  {\bibfnamefont {A.}~\bibnamefont {Assalini}}, \bibinfo {author}
  {\bibfnamefont {K.}~\bibnamefont {Ema}},\ and\ \bibinfo {author}
  {\bibfnamefont {M.}~\bibnamefont {Sasaki}},\ }\href
  {https://doi.org/10.1103/PhysRevA.86.042328} {\bibfield  {journal} {\bibinfo
  {journal} {Phys. Rev. A}\ }\textbf {\bibinfo {volume} {86}},\ \bibinfo
  {pages} {042328} (\bibinfo {year} {2012})}\BibitemShut {NoStop}%
\bibitem [{\citenamefont {Becerra}\ \emph {et~al.}(2013)\citenamefont
  {Becerra}, \citenamefont {Fan}, \citenamefont {Baumgartner}, \citenamefont
  {Goldhar}, \citenamefont {Kosloski},\ and\ \citenamefont
  {Migdall}}]{becerra2013experimental}%
  \BibitemOpen
  \bibfield  {author} {\bibinfo {author} {\bibfnamefont {F.}~\bibnamefont
  {Becerra}}, \bibinfo {author} {\bibfnamefont {J.}~\bibnamefont {Fan}},
  \bibinfo {author} {\bibfnamefont {G.}~\bibnamefont {Baumgartner}}, \bibinfo
  {author} {\bibfnamefont {J.}~\bibnamefont {Goldhar}}, \bibinfo {author}
  {\bibfnamefont {J.}~\bibnamefont {Kosloski}},\ and\ \bibinfo {author}
  {\bibfnamefont {A.}~\bibnamefont {Migdall}},\ }\href@noop {} {\bibfield
  {journal} {\bibinfo  {journal} {Nature Photonics}\ }\textbf {\bibinfo
  {volume} {7}},\ \bibinfo {pages} {147} (\bibinfo {year} {2013})}\BibitemShut
  {NoStop}%
\bibitem [{Note2()}]{Note2}%
  \BibitemOpen
  \bibinfo {note} {For instance, as we saw in \protect \cref {fig:lobes4}(a), a
  one lobe solution is present with $n=3$ with $\beta $ small, implying a small
  effective dimension and consequently, a high storage capacity \cite [sec.
  S6]{SuppMaterial}.}\BibitemShut {Stop}%
\bibitem [{Note3()}]{Note3}%
  \BibitemOpen
  \bibinfo {note} {We note that the Hebbian critical capacity is found in the
  limit of infinite dimension (neurons) and zero temperature while in our case
  the dimension is finite once the Hilbert space is truncated. Nevertheless,
  the critical value seems to apply to finite resources too. For instance,
  \protect \citet {enhancing2021marsh} found numerically that this limit
  persists with at least $200$ neurons, and practical examples never overcome
  this limit \cite {fuchs1988pattern,fiorelli2020signatures}. \label
  {fn:hebbianlimit}}\BibitemShut {NoStop}%
\bibitem [{\citenamefont {Wilson}\ and\ \citenamefont
  {Cowan}(1972)}]{wilson1972excitatory}%
  \BibitemOpen
  \bibfield  {author} {\bibinfo {author} {\bibfnamefont {H.~R.}\ \bibnamefont
  {Wilson}}\ and\ \bibinfo {author} {\bibfnamefont {J.~D.}\ \bibnamefont
  {Cowan}},\ }\href@noop {} {\bibfield  {journal} {\bibinfo  {journal}
  {Biophysical journal}\ }\textbf {\bibinfo {volume} {12}},\ \bibinfo {pages}
  {1} (\bibinfo {year} {1972})}\BibitemShut {NoStop}%
\bibitem [{\citenamefont {Buice}\ and\ \citenamefont
  {Cowan}(2007)}]{PhysRevE.75.051919}%
  \BibitemOpen
  \bibfield  {author} {\bibinfo {author} {\bibfnamefont {M.~A.}\ \bibnamefont
  {Buice}}\ and\ \bibinfo {author} {\bibfnamefont {J.~D.}\ \bibnamefont
  {Cowan}},\ }\href {https://doi.org/10.1103/PhysRevE.75.051919} {\bibfield
  {journal} {\bibinfo  {journal} {Phys. Rev. E}\ }\textbf {\bibinfo {volume}
  {75}},\ \bibinfo {pages} {051919} (\bibinfo {year} {2007})}\BibitemShut
  {NoStop}%
\bibitem [{\citenamefont {Bressloff}(2010)}]{PhysRevE.82.051903}%
  \BibitemOpen
  \bibfield  {author} {\bibinfo {author} {\bibfnamefont {P.~C.}\ \bibnamefont
  {Bressloff}},\ }\href {https://doi.org/10.1103/PhysRevE.82.051903} {\bibfield
   {journal} {\bibinfo  {journal} {Phys. Rev. E}\ }\textbf {\bibinfo {volume}
  {82}},\ \bibinfo {pages} {051903} (\bibinfo {year} {2010})}\BibitemShut
  {NoStop}%
\bibitem [{\citenamefont {Fiorelli}\ \emph {et~al.}(2019)\citenamefont
  {Fiorelli}, \citenamefont {Rotondo}, \citenamefont {Marcuzzi}, \citenamefont
  {Garrahan},\ and\ \citenamefont {Lesanovsky}}]{fiorelli2019quantum}%
  \BibitemOpen
  \bibfield  {author} {\bibinfo {author} {\bibfnamefont {E.}~\bibnamefont
  {Fiorelli}}, \bibinfo {author} {\bibfnamefont {P.}~\bibnamefont {Rotondo}},
  \bibinfo {author} {\bibfnamefont {M.}~\bibnamefont {Marcuzzi}}, \bibinfo
  {author} {\bibfnamefont {J.~P.}\ \bibnamefont {Garrahan}},\ and\ \bibinfo
  {author} {\bibfnamefont {I.}~\bibnamefont {Lesanovsky}},\ }\href@noop {}
  {\bibfield  {journal} {\bibinfo  {journal} {Phys. Rev. A}\ }\textbf {\bibinfo
  {volume} {99}},\ \bibinfo {pages} {032126} (\bibinfo {year}
  {2019})}\BibitemShut {NoStop}%
\bibitem [{\citenamefont {Segura}\ and\ \citenamefont
  {Perazzo}(2000)}]{segura2000associative}%
  \BibitemOpen
  \bibfield  {author} {\bibinfo {author} {\bibfnamefont {E.~C.}\ \bibnamefont
  {Segura}}\ and\ \bibinfo {author} {\bibfnamefont {R.~P.}\ \bibnamefont
  {Perazzo}},\ }\href@noop {} {\bibfield  {journal} {\bibinfo  {journal}
  {Neural Processing Letters}\ }\textbf {\bibinfo {volume} {12}},\ \bibinfo
  {pages} {129} (\bibinfo {year} {2000})}\BibitemShut {NoStop}%
\bibitem [{\citenamefont {Chang}\ \emph {et~al.}(2020)\citenamefont {Chang},
  \citenamefont {Sab\'{\i}n}, \citenamefont {Forn-D\'{\i}az}, \citenamefont
  {Quijandr\'{\i}a}, \citenamefont {Vadiraj}, \citenamefont {Nsanzineza},
  \citenamefont {Johansson},\ and\ \citenamefont {Wilson}}]{forn2020three}%
  \BibitemOpen
  \bibfield  {author} {\bibinfo {author} {\bibfnamefont {C.~W.~S.}\
  \bibnamefont {Chang}}, \bibinfo {author} {\bibfnamefont {C.}~\bibnamefont
  {Sab\'{\i}n}}, \bibinfo {author} {\bibfnamefont {P.}~\bibnamefont
  {Forn-D\'{\i}az}}, \bibinfo {author} {\bibfnamefont {F.}~\bibnamefont
  {Quijandr\'{\i}a}}, \bibinfo {author} {\bibfnamefont {A.~M.}\ \bibnamefont
  {Vadiraj}}, \bibinfo {author} {\bibfnamefont {I.}~\bibnamefont {Nsanzineza}},
  \bibinfo {author} {\bibfnamefont {G.}~\bibnamefont {Johansson}},\ and\
  \bibinfo {author} {\bibfnamefont {C.~M.}\ \bibnamefont {Wilson}},\ }\href
  {https://doi.org/10.1103/PhysRevX.10.011011} {\bibfield  {journal} {\bibinfo
  {journal} {Phys. Rev. X}\ }\textbf {\bibinfo {volume} {10}},\ \bibinfo
  {pages} {011011} (\bibinfo {year} {2020})}\BibitemShut {NoStop}%
\bibitem [{\citenamefont {Haga}\ \emph {et~al.}(2021)\citenamefont {Haga},
  \citenamefont {Nakagawa}, \citenamefont {Hamazaki},\ and\ \citenamefont
  {Ueda}}]{haga2021liouvillian}%
  \BibitemOpen
  \bibfield  {author} {\bibinfo {author} {\bibfnamefont {T.}~\bibnamefont
  {Haga}}, \bibinfo {author} {\bibfnamefont {M.}~\bibnamefont {Nakagawa}},
  \bibinfo {author} {\bibfnamefont {R.}~\bibnamefont {Hamazaki}},\ and\
  \bibinfo {author} {\bibfnamefont {M.}~\bibnamefont {Ueda}},\ }\href@noop {}
  {\bibfield  {journal} {\bibinfo  {journal} {Physical Review Letters}\
  }\textbf {\bibinfo {volume} {127}},\ \bibinfo {pages} {070402} (\bibinfo
  {year} {2021})}\BibitemShut {NoStop}%
\bibitem [{\citenamefont {Mori}(2021)}]{mori2021metastability}%
  \BibitemOpen
  \bibfield  {author} {\bibinfo {author} {\bibfnamefont {T.}~\bibnamefont
  {Mori}},\ }\href@noop {} {\bibfield  {journal} {\bibinfo  {journal} {Physical
  Review Research}\ }\textbf {\bibinfo {volume} {3}},\ \bibinfo {pages}
  {043137} (\bibinfo {year} {2021})}\BibitemShut {NoStop}%
\bibitem [{\citenamefont {Flynn}\ \emph {et~al.}(2021)\citenamefont {Flynn},
  \citenamefont {Cobanera},\ and\ \citenamefont {Viola}}]{flynn2021topology}%
  \BibitemOpen
  \bibfield  {author} {\bibinfo {author} {\bibfnamefont {V.~P.}\ \bibnamefont
  {Flynn}}, \bibinfo {author} {\bibfnamefont {E.}~\bibnamefont {Cobanera}},\
  and\ \bibinfo {author} {\bibfnamefont {L.}~\bibnamefont {Viola}},\
  }\href@noop {} {\bibfield  {journal} {\bibinfo  {journal} {Physical Review
  Letters}\ }\textbf {\bibinfo {volume} {127}},\ \bibinfo {pages} {245701}
  (\bibinfo {year} {2021})}\BibitemShut {NoStop}%
\bibitem [{\citenamefont {Rudolph}\ \emph {et~al.}(2003)\citenamefont
  {Rudolph}, \citenamefont {Spekkens},\ and\ \citenamefont
  {Turner}}]{rudolph2003unambiguous}%
  \BibitemOpen
  \bibfield  {author} {\bibinfo {author} {\bibfnamefont {T.}~\bibnamefont
  {Rudolph}}, \bibinfo {author} {\bibfnamefont {R.~W.}\ \bibnamefont
  {Spekkens}},\ and\ \bibinfo {author} {\bibfnamefont {P.~S.}\ \bibnamefont
  {Turner}},\ }\href {https://doi.org/10.1103/PhysRevA.68.010301} {\bibfield
  {journal} {\bibinfo  {journal} {Phys. Rev. A}\ }\textbf {\bibinfo {volume}
  {68}},\ \bibinfo {pages} {010301(R)} (\bibinfo {year} {2003})}\BibitemShut
  {NoStop}%
\bibitem [{\citenamefont {Azouit}\ \emph {et~al.}(2015)\citenamefont {Azouit},
  \citenamefont {Sarlette},\ and\ \citenamefont
  {Rouchon}}]{azouit2015convergence}%
  \BibitemOpen
  \bibfield  {author} {\bibinfo {author} {\bibfnamefont {R.}~\bibnamefont
  {Azouit}}, \bibinfo {author} {\bibfnamefont {A.}~\bibnamefont {Sarlette}},\
  and\ \bibinfo {author} {\bibfnamefont {P.}~\bibnamefont {Rouchon}},\ }in\
  \href@noop {} {\emph {\bibinfo {booktitle} {2015 54th IEEE Conference on
  Decision and Control (CDC)}}}\ (\bibinfo {organization} {IEEE},\ \bibinfo
  {year} {2015})\ pp.\ \bibinfo {pages} {6447--6453}\BibitemShut {NoStop}%
\bibitem [{\citenamefont {Azouit}\ \emph {et~al.}(2016)\citenamefont {Azouit},
  \citenamefont {Sarlette},\ and\ \citenamefont {Rouchon}}]{azouit2016well}%
  \BibitemOpen
  \bibfield  {author} {\bibinfo {author} {\bibfnamefont {R.}~\bibnamefont
  {Azouit}}, \bibinfo {author} {\bibfnamefont {A.}~\bibnamefont {Sarlette}},\
  and\ \bibinfo {author} {\bibfnamefont {P.}~\bibnamefont {Rouchon}},\
  }\href@noop {} {\bibfield  {journal} {\bibinfo  {journal} {ESAIM: Control,
  Optimisation and Calculus of Variations}\ }\textbf {\bibinfo {volume} {22}},\
  \bibinfo {pages} {1353} (\bibinfo {year} {2016})}\BibitemShut {NoStop}%
\bibitem [{\citenamefont {Chamberland}\ \emph {et~al.}(2022)\citenamefont
  {Chamberland}, \citenamefont {Noh}, \citenamefont {Arrangoiz-Arriola},
  \citenamefont {Campbell}, \citenamefont {Hann}, \citenamefont {Iverson},
  \citenamefont {Putterman}, \citenamefont {Bohdanowicz}, \citenamefont
  {Flammia}, \citenamefont {Keller} \emph {et~al.}}]{chamberland2022building}%
  \BibitemOpen
  \bibfield  {author} {\bibinfo {author} {\bibfnamefont {C.}~\bibnamefont
  {Chamberland}}, \bibinfo {author} {\bibfnamefont {K.}~\bibnamefont {Noh}},
  \bibinfo {author} {\bibfnamefont {P.}~\bibnamefont {Arrangoiz-Arriola}},
  \bibinfo {author} {\bibfnamefont {E.~T.}\ \bibnamefont {Campbell}}, \bibinfo
  {author} {\bibfnamefont {C.~T.}\ \bibnamefont {Hann}}, \bibinfo {author}
  {\bibfnamefont {J.}~\bibnamefont {Iverson}}, \bibinfo {author} {\bibfnamefont
  {H.}~\bibnamefont {Putterman}}, \bibinfo {author} {\bibfnamefont {T.~C.}\
  \bibnamefont {Bohdanowicz}}, \bibinfo {author} {\bibfnamefont {S.~T.}\
  \bibnamefont {Flammia}}, \bibinfo {author} {\bibfnamefont {A.}~\bibnamefont
  {Keller}}, \emph {et~al.},\ }\href@noop {} {\bibfield  {journal} {\bibinfo
  {journal} {PRX Quantum}\ }\textbf {\bibinfo {volume} {3}},\ \bibinfo {pages}
  {010329} (\bibinfo {year} {2022})}\BibitemShut {NoStop}%
\bibitem [{\citenamefont {Meccia}\ and\ \citenamefont
  {Perazzo}(2002)}]{meccia2002biologically}%
  \BibitemOpen
  \bibfield  {author} {\bibinfo {author} {\bibfnamefont {E.~C.~S.}\
  \bibnamefont {Meccia}}\ and\ \bibinfo {author} {\bibfnamefont {R.~P.}\
  \bibnamefont {Perazzo}},\ }\href@noop {} {\bibfield  {journal} {\bibinfo
  {journal} {Neural Processing Letters}\ }\textbf {\bibinfo {volume} {16}},\
  \bibinfo {pages} {243} (\bibinfo {year} {2002})}\BibitemShut {NoStop}%
\bibitem [{\citenamefont {Carmichael}(1999)}]{carmichael1999statistical}%
  \BibitemOpen
  \bibfield  {author} {\bibinfo {author} {\bibfnamefont {H.~J.}\ \bibnamefont
  {Carmichael}},\ }\href@noop {} {\emph {\bibinfo {title} {Statistical methods
  in quantum optics 1: master equations and Fokker-Planck equations}}},\
  Vol.~\bibinfo {volume} {1}\ (\bibinfo  {publisher} {Springer Science \&
  Business Media},\ \bibinfo {year} {1999})\BibitemShut {NoStop}%
\bibitem [{\citenamefont {Johansson}\ \emph {et~al.}(2013)\citenamefont
  {Johansson}, \citenamefont {Nation},\ and\ \citenamefont
  {Nori}}]{Johansson2013qutip}%
  \BibitemOpen
  \bibfield  {author} {\bibinfo {author} {\bibfnamefont {J.}~\bibnamefont
  {Johansson}}, \bibinfo {author} {\bibfnamefont {P.}~\bibnamefont {Nation}},\
  and\ \bibinfo {author} {\bibfnamefont {F.}~\bibnamefont {Nori}},\ }\href
  {https://doi.org/https://doi.org/10.1016/j.cpc.2012.11.019} {\bibfield
  {journal} {\bibinfo  {journal} {Computer Physics Communications}\ }\textbf
  {\bibinfo {volume} {184}},\ \bibinfo {pages} {1234} (\bibinfo {year}
  {2013})}\BibitemShut {NoStop}%
\bibitem [{\citenamefont {Kr{\"a}mer}\ \emph {et~al.}(2018)\citenamefont
  {Kr{\"a}mer}, \citenamefont {Plankensteiner}, \citenamefont {Ostermann},\
  and\ \citenamefont {Ritsch}}]{kramer2018quantumoptics}%
  \BibitemOpen
  \bibfield  {author} {\bibinfo {author} {\bibfnamefont {S.}~\bibnamefont
  {Kr{\"a}mer}}, \bibinfo {author} {\bibfnamefont {D.}~\bibnamefont
  {Plankensteiner}}, \bibinfo {author} {\bibfnamefont {L.}~\bibnamefont
  {Ostermann}},\ and\ \bibinfo {author} {\bibfnamefont {H.}~\bibnamefont
  {Ritsch}},\ }\href@noop {} {\bibfield  {journal} {\bibinfo  {journal}
  {Computer Physics Communications}\ }\textbf {\bibinfo {volume} {227}},\
  \bibinfo {pages} {109} (\bibinfo {year} {2018})}\BibitemShut {NoStop}%
\bibitem [{\citenamefont {Fuchs}\ and\ \citenamefont
  {Haken}(1988)}]{fuchs1988pattern}%
  \BibitemOpen
  \bibfield  {author} {\bibinfo {author} {\bibfnamefont {A.}~\bibnamefont
  {Fuchs}}\ and\ \bibinfo {author} {\bibfnamefont {H.}~\bibnamefont {Haken}},\
  }\href@noop {} {\bibfield  {journal} {\bibinfo  {journal} {Biological
  Cybernetics}\ }\textbf {\bibinfo {volume} {60}},\ \bibinfo {pages} {17}
  (\bibinfo {year} {1988})}\BibitemShut {NoStop}%
\end{thebibliography}%

\clearpage
\onecolumngrid

\appendix
\renewcommand{\thefigure}{S\arabic{figure}}
\def\theequation{S\arabic{equation}}
\renewcommand{\thesection}{S\arabic{section}}
\setcounter{secnumdepth}{2}
\renewcommand\appendixname{}

\begin{center}
    {\Large Supplemental Material}
\end{center}

\section{Mean-field limit} \label{apx:mf}

    To gain intuition about the quantum system, we can analyze the mean-field equations of motion. Hence, we assume $\alpha = \ev{\opa}$ and approximate $\ev{\opa^x (\opad)^y} \sim \ev{\opa}^x\ev{\opad}^y$. Then, we take $\tr \opa \dot{\rho}$ from the master equation \eqref{eq:pp_me_n} and factorize all high-order moments in the right-hand side to obtain
    \begin{equation} \label{eq:pp_mf_a}
        \dot{\alpha} = - \frac{\gamma_1}{2}\alpha -i \Delta \alpha - n \eta (\alpha^*)^{n - 1}e^{-i n\theta} - \frac{m}{2}\gamma_m \abs{\alpha}^{2(m - 1)} \alpha\ ,
    \end{equation}
    which can be written as a coupled ODE system for the real variables $R$ and $\phi$ using $\alpha = R \exp(i \phi)$,
    \begin{align}
        \dot{R} &= -\frac{\gamma_1}{2}R - \frac{m}{2}\gamma_mR^{2m - 1} - n \eta R^{n - 1} \cos[n(\phi-\theta)] \label{eq:pp_mf_R}\\
        \dot{\phi} &= - \Delta + n \eta R^{n - 2} \sin[n(\phi - \theta)] \label{eq:pp_mf_phi}\ .
    \end{align}
    Setting $\dot{R} = \dot{\phi} = 0$, we get the fix point equations of the system which leads to $\sin n(\phi - \theta) = \Delta / n\eta R^{n - 2}$. Thus, a solution can only exist for $\dot{\phi} = 0$ if $\Delta < n \eta R^{n-2}$. In this regime, assuming a large separation between lobes ($R \gg 1$) and small detuning ($\Delta \ll 1$), the phase can be approximated by $\phi - \theta \approx \pi j / n$ with $j=1,\dots,n$. Substituting in the first equation leads to \begin{equation}
        -1 - m \frac{\gamma_m}{\gamma_1} R^{2(m - 1)} - 2n \frac{\eta}{\gamma_1} R^{n - 2} (-1)^j = 0\ ,
    \end{equation}
    where all parameters are positive. Therefore, we need $j$ to be odd for a solution to exist. In this way, again with the assumption $R \gg 1$, we arrive at
    \begin{equation}
        m \gamma_m R^{2(m - 1)} = 2 n \eta R^{n - 2} \Longrightarrow R^{2m - n} = \frac{2n\eta}{m \gamma_m}\ .
    \end{equation}
    The amplitude is identical to the one found in \cref{eq:pp_lobes_coh} for $n = m$. Furthermore, the power of the driving is limited by the power of the nonlinear dissipation, that is $n < 2 m $. For higher values, the amplitude would be inversely proportional to $\eta/\gamma_m$ which can be seen as a reversal between the action of the driving and the dissipation. The case $n = 2m$ has to be treated separately and we obtain a relation $R^{2m - 2} \propto 1 / (2\eta - \gamma_m)$. Thus, the fix points only exist for $2 \eta > \gamma_m$. In the other cases, for $n < 2 m$, we can always find $n$ fix points symmetrically distributed except for $\eta = 0$ where the state $\ket{0}$ is the only stable. This makes sense because all the Lindblad operators are dissipative, i.e. all energy is lost to the environment.

    In the numerical simulations made to obtain the results presented throughout the paper we have chosen $\theta = 0$ for simplicity as its value only shifts the position of the lobes and has no impact on metastability or the lobes. Moreover, we have fixed the detuning to $\Delta = 0.4 \gamma_1$ to situate the system in a regime where lobes can emerge for relatively small amplitudes ($R$). Lastly, since all the parameters are scaled in terms of the linear dissipative rate, we set $\gamma_1 = 1$ in all cases, without affecting the outcomes. The value of the other parameters will be specified in the text when necessary.

\section{Role of linear dissipation}

\begin{figure}[h]
        \centering
        \includegraphics{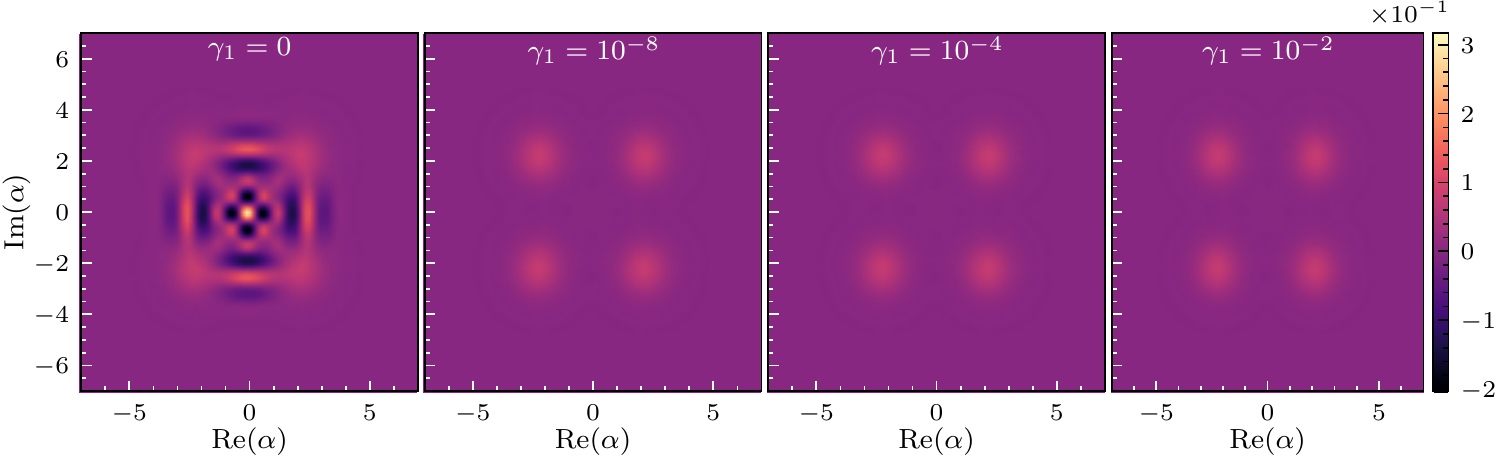}
        \caption{Steady state Wigner representation for different values of $\gamma_1$. Parameters: $n=m=4$, $\gamma_4 = 0.1$, $\eta = 1.14$ and $\Delta = 0.4$.}
        \label{fig:comparison_g1}
    \end{figure}

    In the master equation introduced in the main text, apart from the nonlinear dissipative term shaping the steady state, there is also a linear dissipative term $\disp{\opa}$ with rate $\gamma_1$. The effect of this term is to reduce coherences between the lobes that form the steady state \cite{gevorkyan1999coherent}. We can clearly see this in \cref{fig:comparison_g1}, where we show the Wigner representation for $n=m=4$ and different values of $\gamma_1$. For vanishing $\gamma_1$ the steady state manifold is comprised of four orthogonal $4$-cat states \cite{mirrahimi2014dynamically}, being $\ket{\psi}_{ss} \approx (1/\sqrt{n}) \sum_{j=1}^n \ket{\beta_j}$ an example depicted on the left of \cref{fig:comparison_g1}. Entanglement between the lobes can be appreciated in the negative values acquired by the Wigner representation. Conversely, when $\gamma_1 > 0$, no matter how small, the steady state becomes a statistical mixture of the lobes, i.e. $\rho_{ss} \approx (1/n) \sum_{j=1}^n \op{\beta_j}$.

    Furthermore, the actual value of $\gamma_1$ ($\gamma_1>0$) controls the steady-state decay time. Looking at the $x$ axis of \cref{fig:phase_4} we see that time is normalized in units of this parameter $\tau = \gamma_1 t$. Hence, by reducing $\gamma_1$ we can increase the decay time and, consequently, the metastable time.

    We note that despite working in a finite-dimensional Hilbert space, previous studies have shown that, for vanishing detuning and no linear dissipation, the infinite-dimensional Liouvillian also has $n$ steady states corresponding to the coherent superposition of $n$ symmetrically distributed coherent states (i.e. $n$-cat states) \citep{azouit2016well}. The particular case $n = m = 2$ has also been studied including linear dissipation \citep{azouit2015convergence} with similar results to the ones presented in our work. Hence, even if our study covers a broader range of parameters and regimes where other singular behaviors cannot be excluded, we remark that our truncated solutions provide a good approximation to the analytical ones in infinite Hilbert space, in the mentioned known cases (i.e. $n = m = 2$ or for vanishing detuning and no linear dissipation).

\section{Construction of metastable phases} \label{apx:pp_lobes}
    The fact that the spectrum separates exactly $n$ states allows us to construct $n$ metastable phases as extreme combinations of the corresponding eigenstates \cite{macieszczak2021theory}. We need to be careful as the numerical diagonalization \cite{Johansson2013qutip,kramer2018quantumoptics} returns ``raw'' versions of the eigenvectors $\{\tilde{R}_j, \tilde{L}_j\}$ which are in general not hermitian and do not satisfy the orthogonality conditions between left and right eigenstates. To meet such conditions, we must transform the state within its subspace. Hence, if $\lambda_j \in \mathds{R}$ then $A_j = (\tilde{A}_j + \tilde{A}_j^\dagger)/2$ where $A \in \{L, R\}$ and normalize one of them as $L_j = L_j / \tr\big(L_j^\dagger R_j)$. On the other hand, if $\lambda_j \in \mathds{C}$ then it is guaranteed that exists a second eigenvalue $\lambda_{j + 1} = \lambda_j^*$. So, the corresponding set of eigenstates is $A_j = (A_j + A_{j + 1})/2$ and $A_{j+1} = (A_j - A_{j+1})/2i$ with proper normalization as for the real case.
    
    At this point, Ref.~\cite{macieszczak2021theory} proposed a method to obtain the metastable states as linear combinations of the right eigenvectors. This performs random rotations to the eigenvectors until the volume in coefficient space is maximized. Alternatively, for small $n$, one can find the right combination by using the extreme eigenvalues of the left eigenvectors $(c_j^{min}, c_j^{max})$ \cite{cabot2021metastable}. Then, by inspecting the Wigner form of $R_j$ and using the symmetry of the system we easily find the metastable states for $n = 3$
    \begin{subequations}\label{eq:pp_lobes_RL_3}
    \begin{align}
        \mu_1 &= \rho_{ss} + c_2^{min} R_2, \\
        \mu_2 &= \rho_{ss} + c_2^{max} R_2 + c_3^{min} R_3, \\
        \mu_3 &= \rho_{ss} + c_2^{max} R_2 + c_3^{max} R_3
    \end{align}
    \end{subequations}
    and $n = 4$
    \begin{subequations}\label{eq:pp_lobes_RL_4}
    \begin{align}
        \mu_1 &= \rho_{ss} + c_2^{min} R_2 + c_3^{min} R_3 + c_4^{max} R_4, \\
        \mu_2 &= \rho_{ss} + c_2^{max} R_2 + c_3^{min} R_3 + c_4^{min} R_4, \\
        \mu_3 &= \rho_{ss} + c_2^{max} R_2 + c_3^{max} R_3 + c_4^{max} R_4, \\
        \mu_4 &= \rho_{ss} + c_2^{min} R_2 + c_3^{max} R_3 + c_4^{min} R_4.
    \end{align}
    \end{subequations}
    In general, $c_j^{min} = - c_j^{max}$ and converge to an absolute value of $1$ in the thermodynamic limit. Nevertheless, in the case $n = 3$, we have that $c_2^{min} \approx -2 c_2^{max}$ such that the relation $\rho_{ss} = (1/n)\sum_j \mu_j$ holds.
    
\begin{figure}
        \centering
        \includegraphics[width=\linewidth]{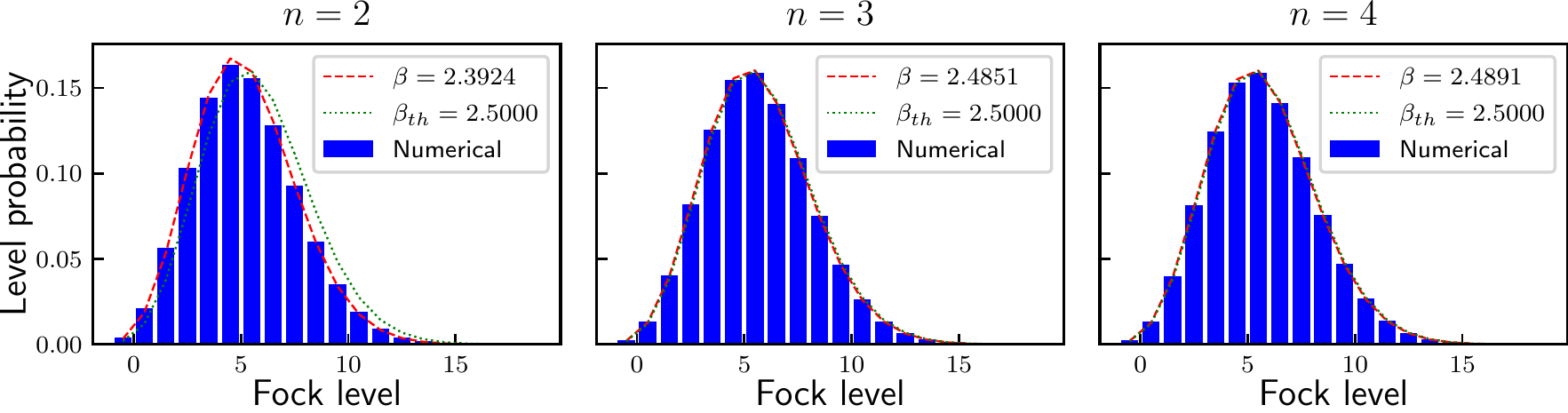}
        \caption{Fock distribution of the steady state for three different values of $n$ with the same $\beta_{th} = 2.5$. Dotted red lines correspond to the numerical value evaluated by fitting the occupation probabilities (blue bars).}
        \label{fig:pp_apx_fock}
    \end{figure}

    In \cref{fig:pp_apx_fock}, one can check that the previous states correspond with high accuracy to the theoretical states in \cref{eq:pp_lobes_coh} of the main text. We can see that as $n$ increases, the numerical $\beta$ gets closer to the theoretical value. Indeed, to obtain \cref{eq:pp_lobes_coh} we neglected the first two terms of \cref{eq:pp_me_n} so when the nonlinear dissipative term gets stronger ($n$ grows), the approximation improves.
    
\section{Liouvillian eigenvalues structure}

    In the previous section, we saw that the relation between the minimum and maximum eigenvalue changes between the even and odd cases. This result can be traced back to the different structures of the Liouvillian spectrum. 
    
\begin{figure}
        \centering
        \includegraphics[width=\linewidth]{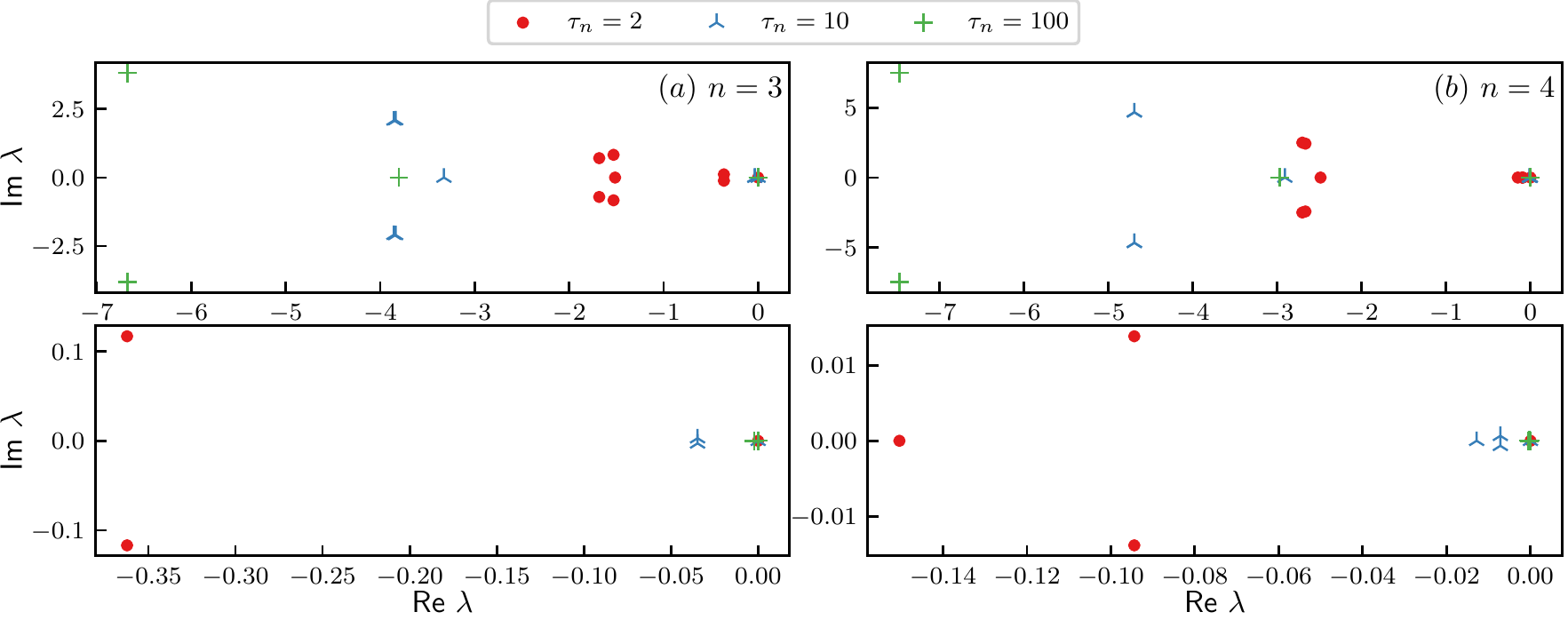}
        \caption{First 9 eigenvalues of the Liouvillian for $n = 3$ (left) and $n = 4$ (right), where we only show those contributing to metastability (the first $n$) in the bottom plots. The parameter sets chosen are the same as in \cref{fig:traj} of the main text.}
        \label{fig:apx_spectra}
    \end{figure}

    In \cref{fig:apx_spectra} we show three different cases for $n = 3$ and $n = 4$ corresponding to the lines in \cref{fig:traj} of the main text. By looking at the global picture (upper row), we appreciate a similar behavior. The slowest eigenmodes are gathered to the right close to the steady state eigenvalue $\lambda_1 = 0$ and as we move to the left (increasing negative real part) we find the fastest eigenmodes that rapidly decay in the time evolution. Of course, the higher $\tau_n$, the larger the separation between slow and fast modes. 

    The most significant difference occurs in the arrangement of the first $n$ modes. On the one hand, for $n = 3$, they form a triangular shape where the second and third eigenvalues are complex conjugates of each other. On the other hand, for $n = 4$, the inclusion of the fourth eigenvalue in the metastable manifold results in a diamond shape with the last eigenvalue ($\tau_4$) being real and slightly separated from $\lambda_{2/3}$. This results in an intermediate time range between the end of the metastable transient ($\tau_4$) and the start of the final decay ($\tau_{2/3}$) which apparently extends the plateau of constant amplitude as seen in \cref{fig:traj}(b) as compared to \cref{fig:traj}(a). Nevertheless, for $t > \tau_4$, the metastable states in \cref{eq:pp_lobes_RL_4} are slowly lost which already reduces the success probability in \cref{fig:memory}. 
    
\section{Ambiguous and Unambiguous POVM}

    In \cref{fig:memory} of the main text, we compare the success probability of identifying the correct lobes between two POVMs: an ambiguous strategy (solid lines) and an unambiguous strategy (dashed lines). Let's see the details for both.

    The ambiguous POVM is obtained numerically from the theory of classical metastability \cite{macieszczak2021theory}. In the ideal case where \cref{eq:pp_lobes_coh} is exact, the projectors consist of a division of the identity around each lobe \cite{lorch2019quantum}, that is
    \begin{align}
        \Id = \frac{1}{\pi} \int d^2\alpha \op{\alpha} = \frac{1}{\pi} \sum_{j=0}^{n-1} \int_{\phi_j - \delta}^{\phi_{j} + \delta} d\varphi \int_0^\infty dR\ R \op{Re^{i\varphi}} = \sum_{j=0}^{n-1} P_j
    \end{align}
    where $\phi_j = \pi (2 j + 1)/n + \theta$ and $\delta = \pi/n$. Performing the integrals gives
    \begin{equation} \label{eq:apx_povm_th}
        P_j = \frac{\Id}{n} - \frac{1}{n} \sum_{k\neq l}^\infty \frac{\Gamma\inpar{\frac{k+l}{2} + 1}}{(k-l)\sqrt{k!l!}}e^{i(\phi_j - \delta) (k - l)}\sin(2\pi \frac{k - l}{n}) \dyad{k}{l}\ .
    \end{equation}
    So all coherent states whose phase is inside $[\phi_j - \delta, \phi_j + \delta]$ will be classified as being in the $j$-th lobe, no matter the absolute amplitude. 
    
\begin{figure}
        \centering
        \includegraphics[width=\linewidth]{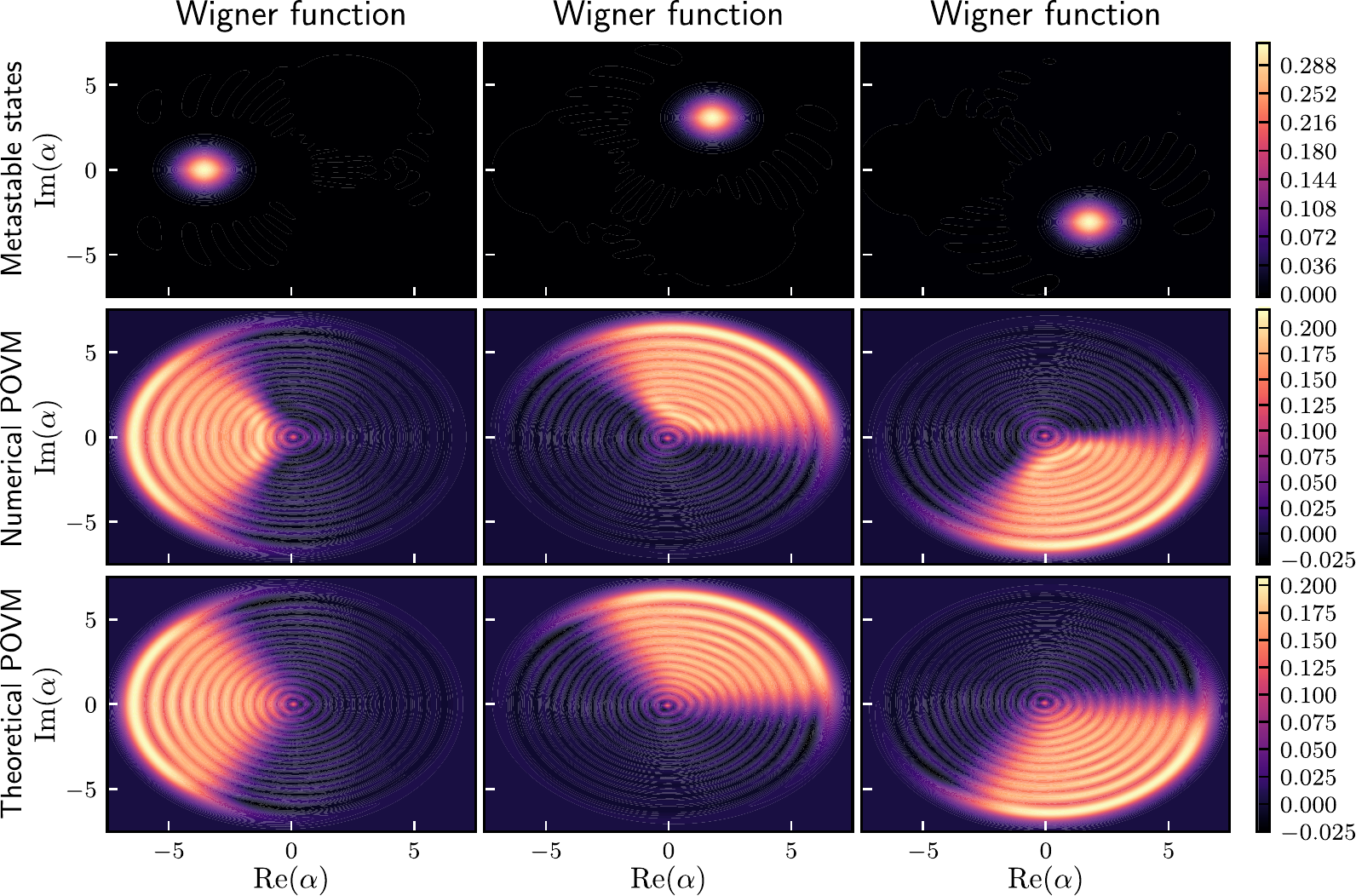}
        \caption{Wigner distribution of the metastable states (top), numerical POVM calculated from \cite{macieszczak2021theory} (middle) and theoretical POVM calculated from \cref{eq:apx_povm_th} (bottom). Parameters: $\gamma_3 = 0.6$, $\eta = 4.6875$, $D = 0.4$, $n = m = 3$.}
        \label{fig:projs_example}
    \end{figure}

    In \cref{fig:projs_example} we compare the theoretical POVM (lower row) obtained from \cref{eq:apx_povm_th} with the projectors obtained numerically (middle row) using the Liouvillian eigenmodes as explained in Ref.~\cite{macieszczak2021theory}. The average trace distance between the two operators is $\sim 0.05$ which results from neglecting the $\Delta$ and $\gamma_1$ terms in \cref{eq:apx_povm_th}. Despite the small difference, in the main text, we have used the numerical POVM to compute the success probability in \cref{fig:memory}(a).
    
    The problem of the previous POVM is the following: our goal is to identify if our system can work as an associative memory, i.e. a state converges to the most similar lobe, but this measure does not permit us to distinguish whether the state is actually over the lobe or in some region nearby.
    
    Therefore, we propose the unambiguous measure which is used for $m$-ary phase-shifted keys \cite{izumi2012displacement,becerra2013experimental}. These types of strategies are characterized by the addition of an extra projector $\Pi_?$ that captures all states different from our targets \cite{rudolph2003unambiguous}, in our case, the lobes. Hence, assuming perfect detector efficiency and no optimization of the displacements, we have $\inbrc{\Pi_j = \op{\beta_j}}_{j=1}^n$ and $\Pi_? = \Id - \sum_{j=1}^n \Pi_j$ which satisfies $\tr \Pi_k \mu_j = \delta_{jk}$ ($\abs{\beta} \gg 1$). An example of the Wigner representation for $\Pi_j$ is given in the first row of \cref{fig:projs_example}. In this way, when the $k$-th operator triggers, we are certain that the state was over the $k$-th lobe. In any other situation, the inconclusive operator $\Pi_?$ will be selected.
    
    In both cases, the probability of measuring the correct lobe $k$ at a time $t$ is $P[\hat{k} \,|\,\rho(t)] = \tr A_{\hat{k}} \rho(t)$ where $A_k \in \{ P_k, \Pi_k \}$. Here, $\hat{k}$ is the estimation of the most similar lobe at time $t = 0$ that is determined using the trace norm, i.e. $\hat{k} = \mathrm{arg min}_{j=1,\dots,n} \norm{\rho(0) - \op{\beta_j}}$. Then, we average $P[\hat{k}\,|\, \rho(t)]$ over $400$ random initial coherent states $\rho(0)$ with $[0, 2 \beta]$ and phase $[0, 2 \pi]$ to avoid biases over particular states. Using this, we obtain the total success probability of identifying the lobe as $P_s = \frac{1}{n} \sum_{k=1}^n \ev*{P[\hat{k}\,|\,\rho(t)]}$. 
    
\section{Critical storage capacity}

    The storage capacity $\alpha_c$ quantifies the density of patterns that can be stored over the size of the system. In the standard \gls{hnn}, the Hebbian rule is used to encode the patterns in the connections between the $N$ spins forming the network. However, it was shown that using this rule only $0.138N$ patterns could be stored and retrieved correctly \cite{amit1985storing}.
    
    In our case, the number of patterns is determined by $n$, the power of the driving, while the size of the system is infinite ($\dim \mathcal{H} = \infty$). However, only the lowest levels of the Fock space are excited once restricted to the metastable manifold \cite{lorch2019quantum}. Therefore, we can truncate the Hilbert space to a level $\dim \mathcal{H}_{eff} = L_{max}$ determining the maximum occupation number that cannot be neglected. Hence, assuming the states in the metastable manifold are coherent and the occupation number follows a Poisson distribution $\{ p_l(\beta)\}$ with mean $\abs{\beta}^2$, we have $L_{max} = \mathrm{arg min}_l \abs{p_l(\beta) - \epsilon}$ where $\epsilon = 10^{-9}$ is an arbitrary numerical accuracy threshold. This leads to the immediate definition of $\alpha_c = n / L_{max}$ which corresponds to the dashed lines in \cref{fig:memory}(b).
    
    This would mean that for small $\beta$ we could have a larger storage capacity compared to the \gls{hnn}. Of course, this is not realistic as for low $\beta$ (small $\eta$) the lobes become indistinguishable, and thus, the success probability becomes $1 / n$. Consequently, we defined a scaled version of $\tilde{\alpha}_c$ that takes into account the separation between lobes, this is
    \begin{equation}
        \tilde{\alpha}_c = \frac{n [1 - F(\beta)]}{L_{max}}\ ,
    \end{equation}
    where $F(\beta) = \abs{\braket{\beta_j}{\beta_{j+1 \bmod n}}}^2$ is the fidelity between two neighbouring lobes that depends on $\beta$ and $n$. We note that as a consequence of the $\mathds{Z}_n$ symmetry, the lobes are identical to each other so the fidelity is equal for all $j=1,\dots,n$. The final result is that for $\beta \to 0 \Rightarrow \alpha_c \to 0$ which correctly describes the impossibility of measuring the lobes independently. 

    In comparing with the $0.138$ Hebbian capacity we would like to mention two points. First, it is the most well-known and established critical value for associative memories. Second, most practical examples that we encountered in the literature store few amount of patterns, even though the system size is large. For instance, \citet{fuchs1988pattern} uses $3600$ classical neurons to store $10$ patterns ($\alpha = 0.0028$) and \citet{fiorelli2020signatures} uses $50$ quantum spins to store $2$ patterns ($\alpha = 0.04$). Further, \citet{enhancing2021marsh} found numerically that this limit persists with at least $200$ neurons.

    We emphasize that in our model, $L_{max}$ corresponds to the effective dimension of the Hilbert space, which is larger than the number of lobes. To see this comparison more clearly, we show in \cref{fig:memory}(c) the minimum size of the system for the required number of patterns. There we can see that for the same amount of patterns, the dimension is smaller in our system than the one required by the Hebbian rule.

    In the previous approach, we choose the Fock basis to calculate the effective dimension of the Hilbert space. It is possible to argue that in the metastable transient the states inside the metastable manifold form a displaced Fock basis spanned by the extreme metastable states \cite{azouit2015convergence,chamberland2022building}. In such a case, the effective dimension of the system is exactly $n$, leading to a storage capacity $\alpha_c = (n / n)[1 - F(\beta)] = 1 - F(\beta)$. Hence, the storage capacity saturates to one for large $\beta$ which is convenient to use as it approaches the ideal limit found by \cite{gardner1988optimal}. Despite that, restricting the dynamics to the metastable manifold limits the choice of initial states as the initial dynamics are not correctly described on such a basis. In this sense, the choice of the Fock basis is the most prudent option, as it allows the dynamics of any initial state to be studied, and despite this, it displays an advantage over the storage capacity of classical systems.
    
\section{Wigner representation}
    
    Using the states, we aim to build a \gls{hnn} \cite{hopfield1982neural} with a single quantum system that can distinguish between the $n$ solutions. The model is dissimilar once compared to the original \gls{hnn} consisting of a net of coupled spins. However, there exist generalizations of the associative network for continuous space \cite{wilson1972excitatory}. 
    In this case, the patterns are excitations of the neuronal field $\inbrc{v_j(x)}_{j=1}^n$. In our system, we can represent the states using the Wigner quasi-probability distribution $\omega(\alpha)$ \cite{carmichael1999statistical} that constitutes a 2d-field. Then, the patterns are the Wigner representations of the lobes in \cref{eq:pp_lobes_coh}, $v_j(\alpha) = \frac{2}{\pi}\exp(-2\abs{\alpha - \beta_j}^2)$. Hence, the evolution of a state is given by
    \begin{equation}
        \pdv{\omega(\alpha; t)}{t} = -\omega(\alpha; t) + g\inpar{\int K(\alpha;\beta)\omega(\beta; t) d^2\beta}
    \end{equation}
    where $g(x)$ is the sigmoid function and the kernel $K(\alpha;\beta) = \sum_{j=1}^n v_j(\alpha)v_j(\beta)$ \cite{meccia2002biologically}. 
    
    The time evolution of the Wigner representation $W(\alpha)$, for $n = m$, is given by
    \begin{align}
         \pdv{W}{t} =& - \gamma_1 W + i \Delta (\alpha\partial_\alpha - {\alpha^*} \partial_{\alpha^*})W  - \frac{\gamma_1}{2}\insqr{\alpha\partial_\alpha + {\alpha^*} \partial_{\alpha^*}  + \partial_\alpha\partial_{\alpha^*}}W \nonumber \\ 
         &+ \frac{\gamma_n}{2}\sum_{k=0}^n \binom{n}{k} \frac{1 - (-1)^k}{2^k} \alpha^{n-k}\partial_{\alpha^*}^k \insqr{\beta^n - \inpar{\alpha^* + \frac{\partial_{\alpha}}{2}}^n}W
    \end{align}
    where $\beta$ has the same definition as in \cref{eq:pp_me_n}.

\end{document}